\documentclass[preprint,10pt,authoryear]{elsarticle}
\usepackage{amsmath}
\usepackage{bm}

\usepackage{booktabs}

\usepackage{adjustbox}
\usepackage{float}
\usepackage{hyperref}
\usepackage{multirow}

\usepackage{subcaption}

\usepackage{amssymb}

\journal{arXiv}

\usepackage{xcolor}

\newcommand{\bpi}{\boldsymbol{\pi}}

\newcommand{\by}{\boldsymbol{y}}
\newcommand{\bY}{\boldsymbol{Y}}

\usepackage{orcidlink}
\usepackage[a4paper, total={7in, 9in}]{geometry}

\usepackage{float}
\usepackage{makecell}
\usepackage{rotating}
\usepackage{longtable}
\usepackage{todonotes}

\usepackage[inline,shortlabels]{enumitem}
\usepackage{ctable}

\usepackage{tcolorbox}
\tcbuselibrary{listings, skins} 
\definecolor{bg}{RGB}{230,230,230} 

\newtcblisting{Rlst}{
	listing only, 
	listing engine=listings,     
	colback=bg, 
	boxsep=2pt, 
	arc=5pt, 
	enhanced, 
	frame hidden,
	listing options={
		language=R,              
		basicstyle=\ttfamily,
		breaklines=true,
		breakindent=1em
	}
}

\begin{document}

\begin{frontmatter}



\title{
A Contaminated  Model for Overdispersed Multinomial Microbiome Count Data}



 \author[inst1]{ Ockert van Heerden \orcidlink{0009-0001-2561-2658}}
 \author[inst1,inst5]{ Andri\"ette Bekker \orcidlink{0000-0003-4793-5674}}
 \author[inst1]{Seite Makgai \orcidlink{0000-0003-3533-9576}}
 \author[inst1]{Arno Otto \corref{cor1}\orcidlink{0000-0002-6565-2910}}
 \ead{arno.otto@up.ac.za}
 \cortext[cor1]{Corresponding author}
 \author[inst2]{Antonio Punzo \orcidlink{0000-0001-7742-1821}}

 \affiliation[inst1]{organization={Department of Statistics, University of Pretoria, Pretoria, South Africa}}
 \affiliation[inst5]{organization= {National Institute for Theoretical and Computational Sciences (NITheCS), Pretoria Node, University of Pretoria, South Africa}}
 \affiliation[inst2]{organization={Department of Economics and Business, University of Catania, Catania, Italy}}


\begin{abstract}
Multinomial count data, such as microbial composition profiles derived from sequencing studies, frequently contain anomalous observations that distort parameter estimates. The Dirichlet-multinomial (DM) distribution is widely used in this setting but remains sensitive to such contamination. We propose the contaminated Dirichlet-multinomial (CDM) distribution, a two-component mixture in which the regular data come from a DM component with a lower dispersion and the irregular data come from a DM component with an inflated dispersion parameter. This construction accommodates anomalies without requiring their removal, and yields a natural rule for anomaly detection via posterior probabilities. Through sensitivity analyses involving both single-point anomalies and background noise, we demonstrate that the CDM distribution effectively downweights the influence of anomalous observations on the parameter estimates. 
The model is applied to gut microbiome data from a colorectal carcinogenesis study, where it consistently outperforms the DM distribution across all information criteria and identifies biologically plausible anomaly proportions in both the healthy and carcinoma subsets. 
\end{abstract}

\begin{keyword}
count data \sep contamination \sep Dirichlet-multinomial \sep outliers \sep overdispersion
\end{keyword}

\end{frontmatter}

\section{Introduction} 
\label{sec:Intro}

The analysis of multinomial count data is a fundamental challenge across a wide array of disciplines, ranging from genomic sequencing to natural language processing and consumer choice modelling. 
One particularly consequential setting is the human gut microbiome, which constitutes one of the most complex ecological systems in human biology. 
It comprises trillions of microbial organisms that maintain a dynamic relationship with host physiology. 
Disruptions to this microbial equilibrium---a phenomenon known as dysbiosis---have been increasingly implicated in the pathogenesis of a wide range of chronic and malignant diseases, making the statistical characterization of gut microbial communities an important priority in biomedical research.

The multinomial distribution is the canonical model for such data, but it imposes a restrictive mean–variance relationship: given the total count $m \in \mathbb{N}_+$, the variance of each category is entirely determined by its mean. 
This assumption is often violated in practice, where the observed variability substantially exceeds that predicted by the multinomial distribution.

A natural and widely adopted solution is the Dirichlet–multinomial (DM) distribution \citep{Mosimann_1962}. 
This distribution extends the multinomial through a compound construction, whereby the underlying category probability vector follows a Dirichlet distribution. 
By accommodating the heterogeneity inherent in count data---namely, the excess variation that the multinomial model fails to capture---the DM has seen widespread adoption in fields such as microbiome composition analysis \citep{Holmes_2012, chen2013variable, wadsworth2017integrative, harrison2020dirichlet}, ecological cover data modelling \citep{damgaard2018joint}, and discrete choice modelling \citep{guimaraes2005dirichlet}.

The traditional parameterisation of the DM distribution may not be immediately intuitive. 
Each parameter simultaneously influences both the expected proportions and the variance, making it difficult to directly interpret how changes in a parameter affect the distributional behaviour. This lack of interpretability poses a barrier for applied researchers who require parameters with a clear real-world meaning.
An alternative parametrization has been proposed to address this limitation by \citet{tvedebrink2009overdispersion}, where the DM distribution was expressed in terms of population allele frequencies.
A mode-based parametrisation is also possible---\citet{Tomarchio_2020} exploit the mode of the Dirichlet distribution for this purpose---although this approach defines a restricted subclass of the DM distribution, resulting in a considerable loss of flexibility. 


From a statistical perspective, the mean plays a central role as it represents the primary quantity of interest in many applications and often admits a direct real-world interpretation. 
In this spirit of the mean parameterised beta-binomial distribution \cite{otto2026modeling}, a Dirichlet–multinomial (DM) distribution in terms of its mean is also followed, yielding a more interpretable formulation. We refer to this as the mean-parameterised DM distribution.

The probability mass function (PMF) of the mean-parameterised DM distribution for a random vector $\bY = (Y_1, \ldots, Y_D)^\top$ with support $\mathbb{S}_m^D=\left\{ \by \in \mathbb{N}_0^D : \sum_{d=1}^D y_d = m \right\}$ is given by
\begin{equation}
    \begin{aligned}
        f_{\text{DM}_m}({\by};\bpi,\sigma) & = \frac{m!}{y_{1}! y_{2}! \cdots y_{D}!} 
        \frac{\Gamma\left( \frac{1}{\sigma} \right)}{\Gamma\left(  m + \frac{1}{\sigma} \right)} 
        \prod_{d=1}^{D}\frac{ \Gamma\left(y_{d}+\frac{\pi_d}{\sigma}\right) }{ \Gamma\left(\frac{\pi_d}{\sigma}\right)},
        \quad \by \in \mathbb{S}_m^D \\ 
    \end{aligned}     \label{eq:DM_PPV}
\end{equation}
where $\bpi = (\pi_1, \ldots, \pi_D)^\top$ is the vector of mean proportions satisfying $\pi_d \in (0,1)$ with $\sum_{d=1}^{D} \pi_d = 1$, $\sigma >0$ is the dispersion parameter, $m = \sum_{d=1}^{D} y_d$ is the fixed total count, and $\Gamma(\cdot)$ denotes the gamma function. 
Although there are $D$ mean proportion parameters, only $D-1$ are free, since they are strictly positive and sum to one. 
Hence, one parameter is redundant and can be expressed as a function of the others.
A similar constraint applies to the entries of $\bY$: although there are $D$ components, only $D-1$ are free, since they are non-negative integers satisfying $\sum_{d=1}^D Y_d = m$.  
If $\bY$ has PMF \eqref{eq:DM_PPV}, we write $\bY \sim \mathcal{DM}_m(\bpi, \sigma)$.
 The components of the expectation, variance, and covariance of $\bY\sim \mathcal{DM}_m(\bpi, \sigma)$ are given by
         \begin{align}
            \text{E}_{\text{DM}_m}(Y_d) & = m \pi_d = \text{E}_{\text{M}_m}(Y_d), \label{eq:DM_mean}\\ 
            \text{Var}_{\text{DM}_m}(Y_d) & = m \pi_d (1-\pi_d) \frac{1+m\sigma}{1+\sigma} \nonumber\\ 
            &= \text{Var}_{\text{M}_m}(Y_d)\frac{1+m\sigma}{1+\sigma}, \label{eq:DM_var}\\ 
            \text{Cov}_{\text{DM}_m}(Y_d,Y_j) & = -m \pi_d \pi_j \frac{1+m\sigma}{1+\sigma} 
            \nonumber \\ 
            &= \text{Cov}_{\text{M}_m}(Y_d,Y_j)\frac{1+m\sigma}{1+\sigma}, \label{eq:DM_cov}
        \end{align}
where $d,j \in \{1,\ldots,D\}$, $d \neq j$, and the quantities with subscript $\text{M}_m$ refer to a multinomial distribution with the same total count $m$.
Hence, the parameterisation in \eqref{eq:DM_PPV} allows for a direct interpretation of the mean structure of the DM distribution.
In particular, a comparison with the multinomial distribution highlights the role of the additional parameter $\sigma$.
While the expected value in \eqref{eq:DM_mean} remains unchanged, both the variance in \eqref{eq:DM_var} and the covariance in \eqref{eq:DM_cov} are inflated by the common factor $\frac{1+m\sigma}{1+\sigma}$.
In contrast to the multinomial distribution, where the variance and covariance are completely determined by the mean proportions, the DM distribution allows the level of dispersion to be adjusted through $\sigma$.
Moreover, the multiplicative form of the inflation factor implies that overdispersion is homogeneous across all components and pairwise covariances.
As $\sigma \to 0$, the inflation factor approaches one and the DM distribution reduces to the multinomial model, whereas increasing values of $\sigma$ lead to progressively larger deviations from the multinomial variance–covariance structure.

Although the DM distribution relaxes the highly restrictive variance assumptions of the multinomial model, there are situations in which it remains insufficiently flexible to capture the true variability in the data. Similar limitations have been noted in other count models \citep{Otto_2025_NB, otto2026modeling}.
While standard overdispersion refers to variability exceeding that of the multinomial model, further overdispersion relative to the DM model may arise when the observed variance exceeds the level accommodated by the DM distribution. Consequently, the DM distribution may be inadequate for modelling the extreme variability present in certain datasets.




In bounded count data, one common cause of such excess variation is an additional data-generating process, such as the presence of an excess of observations near the boundaries. In microbiome data, a further major contributor to inflated variance is the presence of values that are extreme relative to a reference distribution (here assumed to be the DM; see \citet{davies1993identification} and \citet{hennig2002fixed} for a discussion of this concept). 
For simplicity and consistency, we refer to such observations as anomalous observations throughout the paper. 
As discussed by \citet{ritter2014robust}, real-world data is often contaminated with anomalous observations that disproportionately affect the estimation of model parameters. 
The inappropriate imposition of the DM model in these scenarios could ultimately lead to misleading biological inferences.

This raises the question of how anomalies should be handled in multinomial count data. 
To address this, it is useful to distinguish between two broad types of anomalies \citep{ritter2014robust}. 
Gross anomalies are unpredictable observations that cannot be adequately described by any probabilistic model.
In the presence of gross anomalies, the recommended approach is to remove the observations entirely or to adopt a method specifically designed to suppress their effect \citep{barnett1994outliers}.
Mild anomalies, instead, correspond to observations sampled from a population that differs from the assumed model. 
 For mild anomalies, it is generally preferable to employ a model that is sufficiently flexible to accommodate such atypical observations \citep{ritter2014robust,punzo2016parsimonious}, which is the main focus of this paper.

To this end, we propose the contaminated Dirichlet–multinomial (CDM) model, in which observations are assumed to arise from a reference DM distribution, while a proportion of the data (the contamination) is generated from a DM distribution with higher dispersion. 
This formulation follows the contaminated-model framework commonly adopted in the literature (e.g., \citealp{mazza2020mixtures,punzo2017robust,punzo2021multiple}), where anomalous observations are defined in terms of deviations from a reference model rather than their relative frequency.
The resulting model takes the form of a two-component mixture: one component represents the typical observations (reference DM distribution), while the other, sharing the same mean but with an inflated dispersion parameter, represents mild anomalies (contaminant distribution), also referred to as “bad” observations in the nomenclature of \citet{aitkin1980mixture}.
This structure provides sufficient flexibility to accommodate mild anomalies while maintaining parsimony, as both components share a common mean.


The paper is structured as follows. 
Section~\ref{sec:Methodology} introduces the proposed CDM distribution, addresses its identifiability, and presents the corresponding maximum likelihood estimation, along with a discussion on parameter initialization.
Section~\ref{sec:Simulation Study} presents two sensitivity analyses---based on single-point anomalies and background noise---to investigate the impact of anomalous observations on parameter estimation under different scenarios. 
In Section~\ref{sec:Data Application}, we demonstrate the effectiveness of the CDM model as an alternative for modelling overdispersed data, using phylum-level bacterial counts (Firmicutes, Proteobacteria, Bacteroidetes, Fusobacteria, Actinobacteria, and an aggregated remainder) from the healthy and carcinoma subsets of the colorectal cancer microbiome dataset of \cite{nakatsu2015gut}. 
Finally, conclusions are drawn in Section~\ref{section:conclusion}, while a practitioner’s guide to implementing the methodology using an \textsf{R} package is provided in the appendix.

\section{Methodology}
\label{sec:Methodology}

In this section, we introduce the CDM distribution (Section~\ref{Section contaminated dirichlet multinomial}), and outline the corresponding maximum likelihood (ML) estimation of the parameters (Section~\ref{subsec:MLE}). 

\subsection{Contaminated Dirichlet-Multinomial distribution} 
\label{Section contaminated dirichlet multinomial}

The PMF of the proposed CDM distribution is
\begin{align}\label{eq:CDM_PMF}
        f_{\text{CDM}_m}(\by;\bpi,\sigma,\delta,\eta) & = (1-\delta) \underbrace{f_{\text{DM}_m}(\by;\bpi,\sigma)}_{\text{reference }} + \delta \underbrace{f_{\text{DM}_m}(\by;\bpi,\eta \sigma)}_{\text{contaminant}},\quad \by\in \mathbb{S}_m^D ,
\end{align}
where $\delta\in(0,1)$ and $\eta>1$. 
If $\bY$ follows the distribution in \eqref{eq:CDM_PMF}, we denote it as $\bY\sim \mathcal{CDM}_m(\bpi,\sigma,\delta,\eta)$. 
The contamination parameters $\delta$ and $\eta$ have practical interpretations: $\delta$ is the proportion of points not from the reference distribution, while $\eta$ represents the degree of contamination and, because of the assumption $\eta > 1$, it can be viewed as an inflation parameter, i.e., the increase in variability due to the points that do not come from the reference DM distribution.


The coordinates of the expected value, variance, and covariance of $\bY\sim \mathcal{CDM}_m(\bpi,\sigma,\delta,\eta)$ are,
    \begin{align}        
    \text{E}_{\text{CDM}_m}(Y_d) & = m \pi_d , \label{eq:CDM_mean_moments}\\    \text{Var}_{\text{CDM}_m}(Y_d) & = m \pi_d (1-\pi_d) \left[(1-\delta)\frac{1+m\sigma}{1+\sigma}+\delta\frac{1+m\eta\sigma}{1+\eta\sigma}\right] \nonumber \\ 
    &=\text{Var}_{\text{DM}_m}(Y_d)[1+\delta(\kappa_m(\sigma,\eta)-1)], \label{eq:CDM_Var}\\
    \text{Cov}_{\text{CDM}_m}(Y_d,Y_j) & = -m \pi_d \pi_j \left[(1-\delta)\frac{1+m\sigma}{1+\sigma}+\delta\frac{1+m\eta\sigma}{1+\eta\sigma}\right]  \nonumber\\         
    & = \text{Cov}_{\text{DM}_m}(Y_d,Y_j)[1+\delta(\kappa_m(\sigma,\eta)-1)],\label{eq:CDM_Cov}
    \end{align} 
where $d,j \in \{1,\ldots,D\}$, $d\neq j$, and $\kappa_m(\sigma,\eta)=\frac{(1+m\eta\sigma)(1+\sigma)}{(1+\eta\sigma)(1+m\sigma)}$. 
Since $\sigma>0$ and $\eta > 1$, it follows that $\kappa_m(\sigma,\eta)>1$ if $m>1$. 
Consequently, since $\delta \in (0,1)$, the term $[1+\delta(\kappa_m(\sigma,\eta)-1)] > 1$ in \eqref{eq:CDM_Var} and \eqref{eq:CDM_Cov} can be interpreted as an inflation factor for the overall variance and covariance. 
The overall variance and covariance in \eqref{eq:CDM_mean_moments} exceed those of the $\mathcal{DM}_m(\bpi,\sigma)$ distribution, with the extent of the increase determined by the values of $\delta$ and $\eta$. 
Therefore, the CDM distribution can accommodate possible DM overdispersion. When $D=2$ in \eqref{eq:CDM_PMF}, the $\mathcal{CDM}_m(\bpi,\sigma,\delta,\eta)$ simplifies to the contaminated beta-binomial distribution proposed in \citet{otto2026modeling}.

\subsection{Identifiability}
\label{subsec:Identifiability}

Identifiability is a fundamental prerequisite for statistical inference, ensuring that distinct parameter values correspond to distinct probability distributions. 
This property underlies the consistency and asymptotic normality of the maximum likelihood (ML) estimators that we will discuss in Section~\ref{subsec:MLE}. 
For finite mixtures, identifiability is typically defined in terms of the underlying mixing distribution, so that equality of mixture densities implies equality of the corresponding mixing measures, up to label switching \citep{teicher1963identifiability,titterington1985statistical}. 
In the context of Dirichlet-based models, identifiability has recently been studied by \citet{nguyen2026identifiability}.

As discussed in Section \ref{Section contaminated dirichlet multinomial}, the proposed CDM model can be written as a two-component DM mixture consisting of: 
(i) a reference DM distribution with parameters $(\bpi,\sigma)$, and 
(ii) a contaminant DM distribution sharing the same mean vector $\bpi$ and having inflated dispersion $\eta\sigma$, with $\eta > 1$. 
The corresponding mixing proportion is governed by $\delta \in (0,1)$. 
Since the CDM model is a finite mixture, identifiability must be considered with respect to the associated mixing distribution, which in this case is a two-point discrete measure supported on $(\bpi,\sigma)$ and $(\bpi,\eta\sigma)$, with masses $1-\delta$ and $\delta$, respectively. 
In general, mixtures of Dirichlet-type distributions are not identifiable on the unrestricted parameter space. 
However, \citet[][Section~4]{nguyen2026identifiability} show that identifiability is recovered when the number of mixture components is strictly less than the dimension $D$ of the simplex. 
Because the CDM model involves only two components, this condition is satisfied whenever $D \geq 3$, ensuring identifiability in the sense of uniqueness of the mixing measure.
The case $D=2$ (corresponding to the beta-binomial setting) requires separate consideration. 
In this setting, unrestricted DM mixtures are known to be non-identifiable \citep{nguyen2026identifiability}. However, these negative results do not automatically extend to the CDM model, since the latter imposes a common-mean constraint across mixture components. 
Whether this restriction is sufficient to restore identifiability in the two-dimensional case remains an open question and would require a dedicated analysis.

Beyond the general mixture result, the specific structure of the CDM model helps eliminate further sources of ambiguity. 
First, the constraint $\eta > 1$ ensures that the two components have distinct dispersion parameters, thereby preventing label switching and inducing an intrinsic ordering between the components, with the contaminant component being more dispersed than the reference component. 
Second, the assumption $\delta \in (0,1)$ excludes degenerate cases in which one of the components vanishes.

Taken together, these results imply that, for $D \geq 3$, the parameter vector $(\bpi,\sigma,\delta,\eta)$ associated with the CDM model is identifiable from the distribution. 
This provides a rigorous foundation for ML estimation and subsequent statistical inference.

\subsection{Maximum likelihood parameter estimation, model comparison, and outlier detection} 
\label{subsec:MLE}

As briefly discussed in Section~\ref{subsec:Identifiability}, parameter estimation for the CDM distribution in \eqref{eq:CDM_PMF} is carried out via maximum likelihood (ML). 
Given a random sample $\{\by_i\}_{i=1}^n$ of size $n$ from the PMF in \eqref{eq:CDM_PMF}, the corresponding log-likelihood function is
\begin{align}\label{eq:loglike}
\ell(\bpi,\sigma,\delta,\eta)=\sum_{i=1}^n \log\bigl[f_{\text{CDM}_m}(\by_i;\bpi,\sigma,\delta,\eta)\bigr].
\end{align}
We numerically maximise \eqref{eq:loglike} with respect to $\bpi$, $\sigma$, $\delta$, and $\eta$ using the general-purpose optimizer \texttt{nlm()} in \textsf{R} \citep{R_core}, as implemented in the \textbf{stats} package.
We recommend first fitting the DM distribution and then using the resulting estimates as starting values for the ML estimation of the CDM model. 
For $\delta$ and $\eta$, we suggest selecting initial values close to the DM limit, namely $\delta \to 0^+$ and $\eta \to 1^+$.

For model comparison among models whose parameters are estimated via maximum likelihood, we use the Akaike information criterion (AIC; \citealt{Akaike_1974}),
\begin{equation} \label{AIC}
\text{AIC}(\boldsymbol{\hat{\kappa}}) = -2\ell(\boldsymbol{\hat{\kappa}}) + 2\psi,
\end{equation}
and the Bayesian information criterion (BIC; \citealt{Schwarz_1978}),
\begin{equation} \label{BIC}
\text{BIC}(\boldsymbol{\hat{\kappa}}) = -2\ell(\boldsymbol{\hat{\kappa}}) + \psi \log n,
\end{equation}
where $\psi$ denotes the number of free parameters in the model and $\boldsymbol{\hat{\kappa}}$ is the vector of ML estimates. 
For both criteria, smaller values indicate a better model fit. 
For the CDM model, the total number of free parameters is $\psi = D + 2$.

An advantage of model \eqref{eq:CDM_PMF} is that, given the (ML) parameter estimates $\hat\bpi$, $\hat\sigma$, $\hat\delta$, and $\hat\eta$, we can assess whether a generic data point $\by$ is anomalous via the \emph{a posteriori} probability
\begin{equation}
\label{eq:posterior_prob}
P\bigl(\by \text{ arises from } \mathcal{DM}_m(\bpi,\sigma) \mid \hat\bpi, \hat\sigma, \hat\delta, \hat\eta \bigr)
=
\frac{(1-\hat\delta)\, f_{\text{DM}_m}(\by;\hat\bpi,\hat\sigma)}
{f_{\text{CDM}_m}(\by;\hat\bpi,\hat\sigma,\hat\delta,\hat\eta)}.
\end{equation}
Based on \eqref{eq:posterior_prob}, $\by$ is classified as a typical observation if such a probability is greater than 0.5, and as an anomalous observation otherwise. As discussed in \citet{melnykov2026use}, this threshold can be tuned to reflect the practitioner's tolerance for misclassification errors; for instance, a lower threshold may be preferred in applications where failing to flag an anomalous observation carries greater consequences than incorrectly labeling a typical one as anomalous.

\section{Simulation Study}
\label{sec:Simulation Study}

In this section, we perform two sensitivity analyses that investigate the ability of the  CDM distribution to account for anomalies. Specifically, in Section \ref{subsec:location} we evaluate the impact of a single anomaly on the parameter estimates of the DM and CDM distributions. 
While in Section \ref{subsec:variance} we evaluate the impact of background noise. Furthermore, the ability of the CDM model to automatically detect mild anomalies is also evaluated using the true positive rate (TPR), which measures the proportion of anomalous observations that are correctly identified as anomalies, and the false positive rate (FPR), which corresponds to the proportion of typical points incorrectly classified as anomalies.


\subsection{Sensitivity analysis: Impact of a single anomalous observation} 
\label{subsec:location}

\begin{figure}[!ht] 
	    \centering
	    \begin{subfigure}{.5\textwidth}
		    \centering
            \includegraphics[scale=0.55]{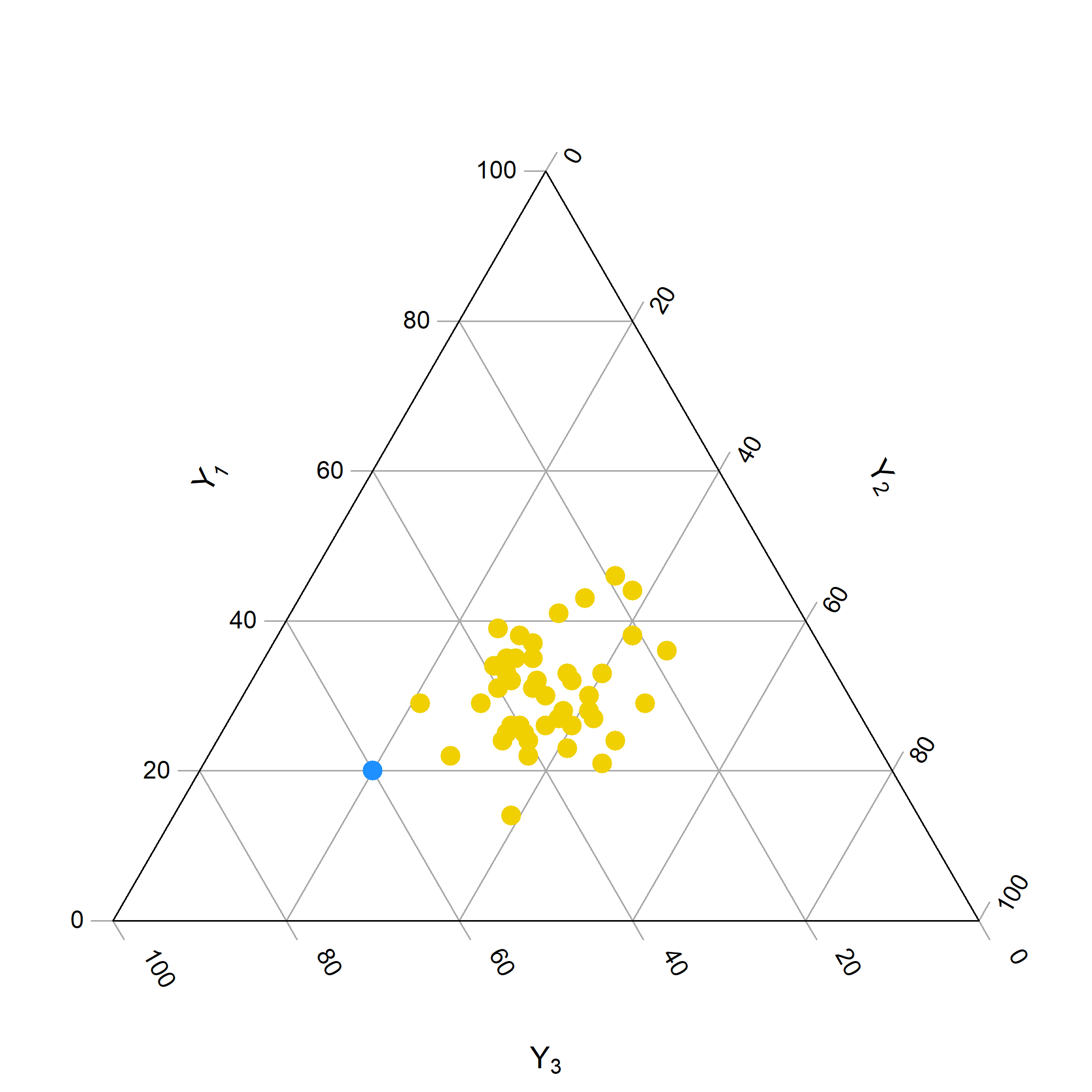}  \label{fig:L_C1}
            \subcaption{Location A: $\by=(20,20,60)$}
	    \end{subfigure}%
        \begin{subfigure}{.5\textwidth} 
		    \centering
            \includegraphics[scale=0.55]{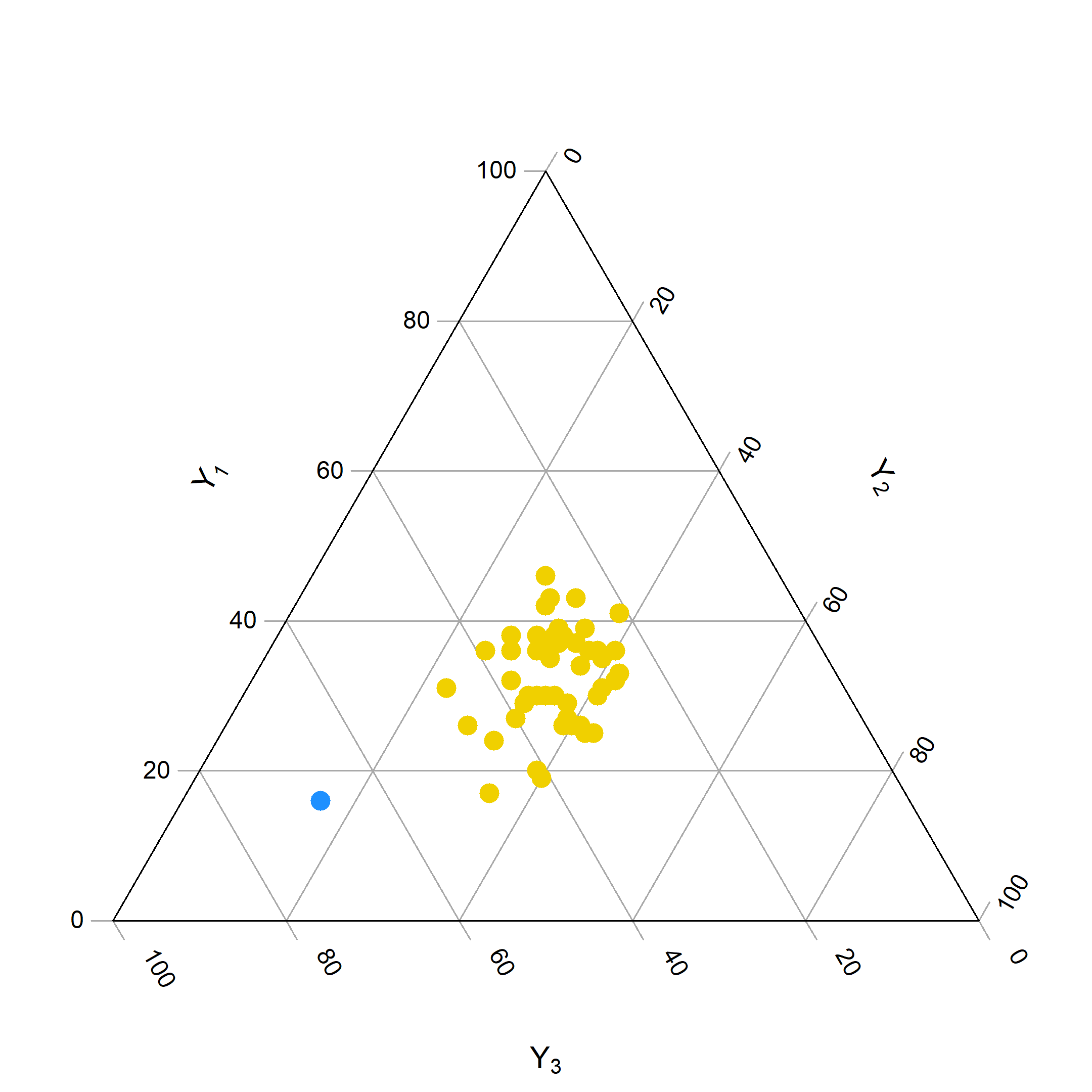}  \label{fig:L_C2}
            \subcaption{Location B: $\by=(16,16,68)$}
	    \end{subfigure}\\
        \begin{subfigure}{.5\textwidth}
		    \centering
            \includegraphics[scale=0.55]{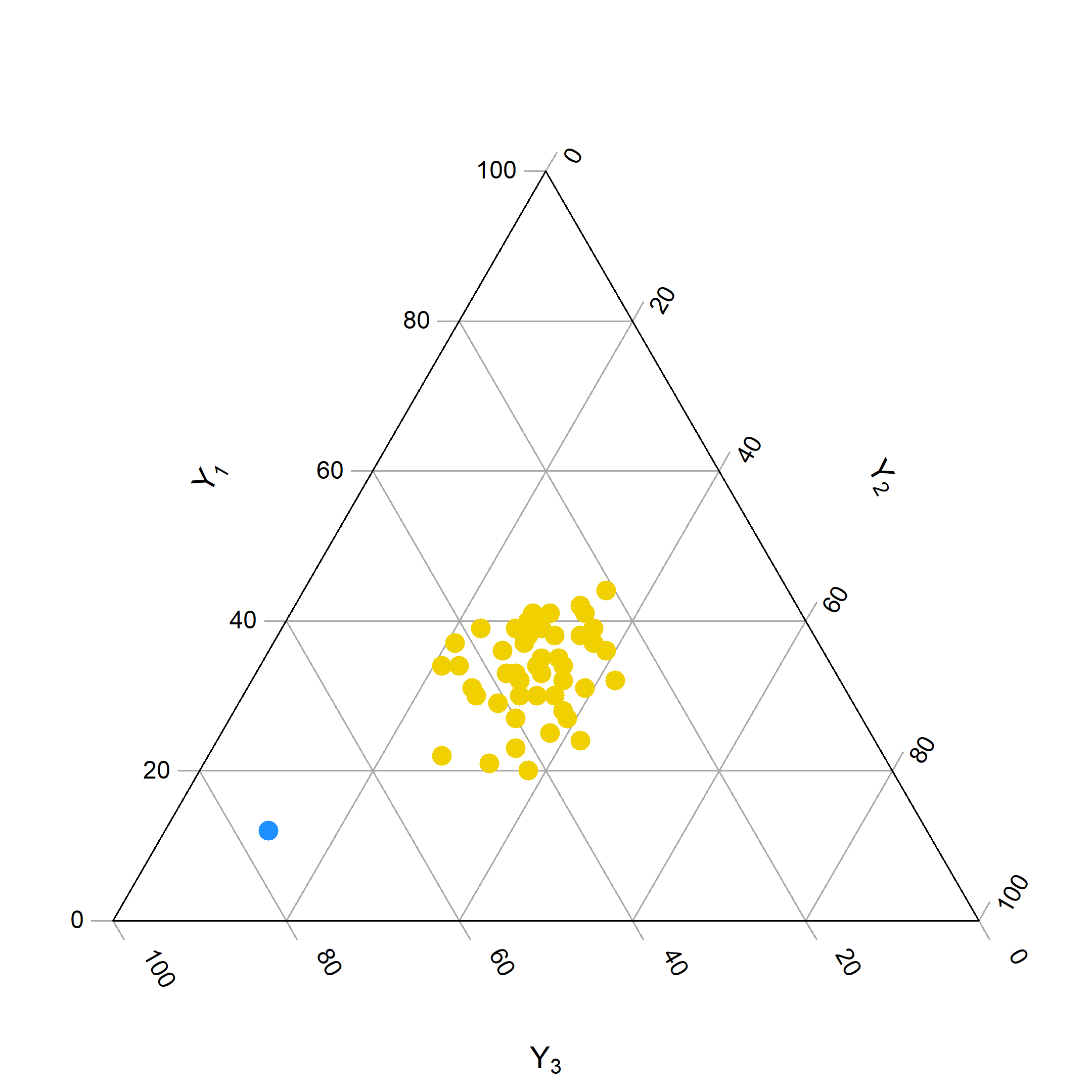}  \label{fig:L_C3}
            \subcaption{Location C: $\by=(12,12,76)$}
	    \end{subfigure}%
        \begin{subfigure}{.5\textwidth} 
		    \centering
            \includegraphics[scale=0.55]{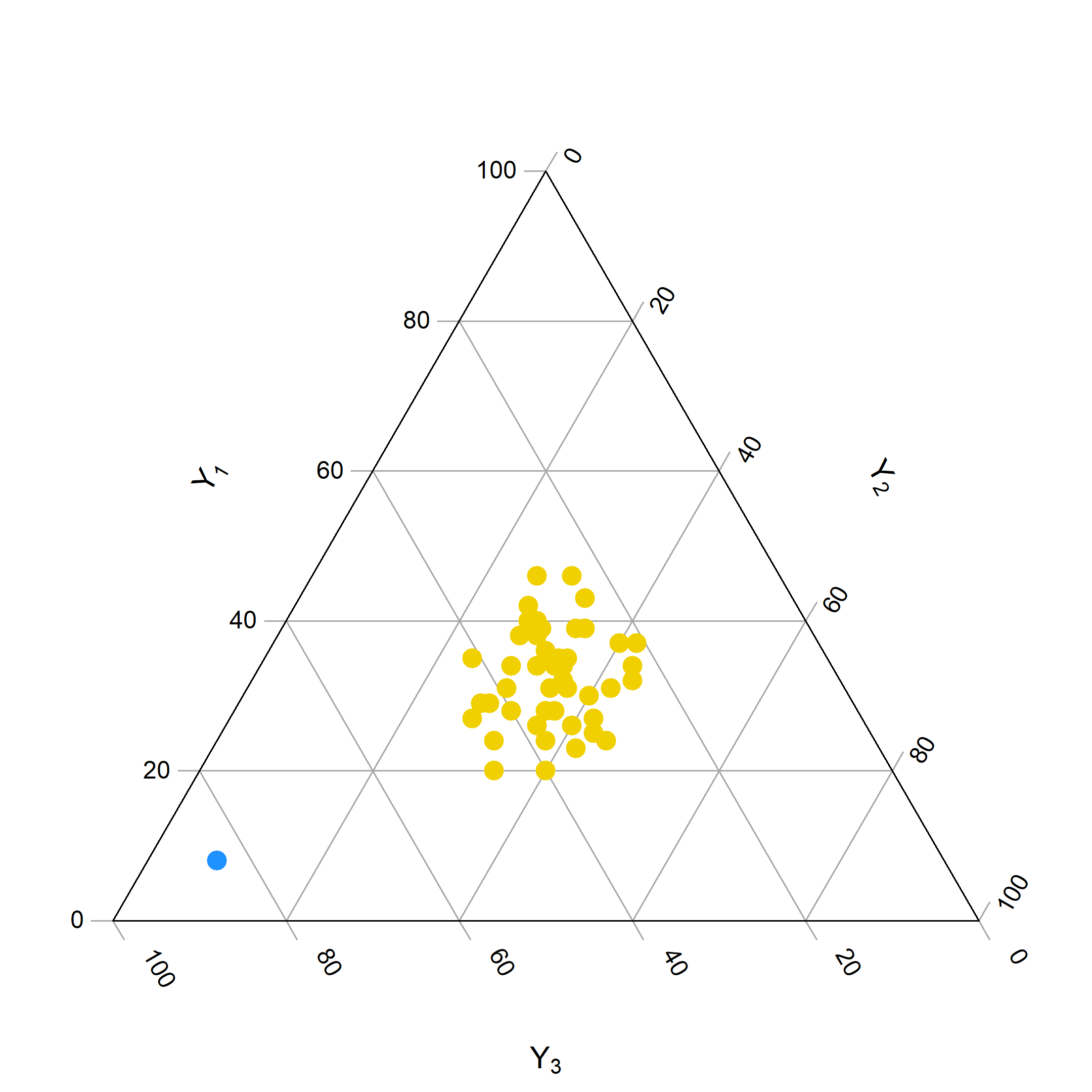}  \label{fig:L_C4}
            \subcaption{Location D: $\by=(8,8,84)$}
	    \end{subfigure}%
        \caption{Four simulated datasets from Scenario 1 of the sensitivity analysis (impact of a single anomalous observation), corresponding to the four anomaly locations (A--D). 
        The anomalous point is shown in blue.}
        \label{fig:L_C} 
    \end{figure}

\begin{figure}[!ht] 
	    \centering
	    \begin{subfigure}{.5\textwidth}
		    \centering
            \includegraphics[scale=0.55]{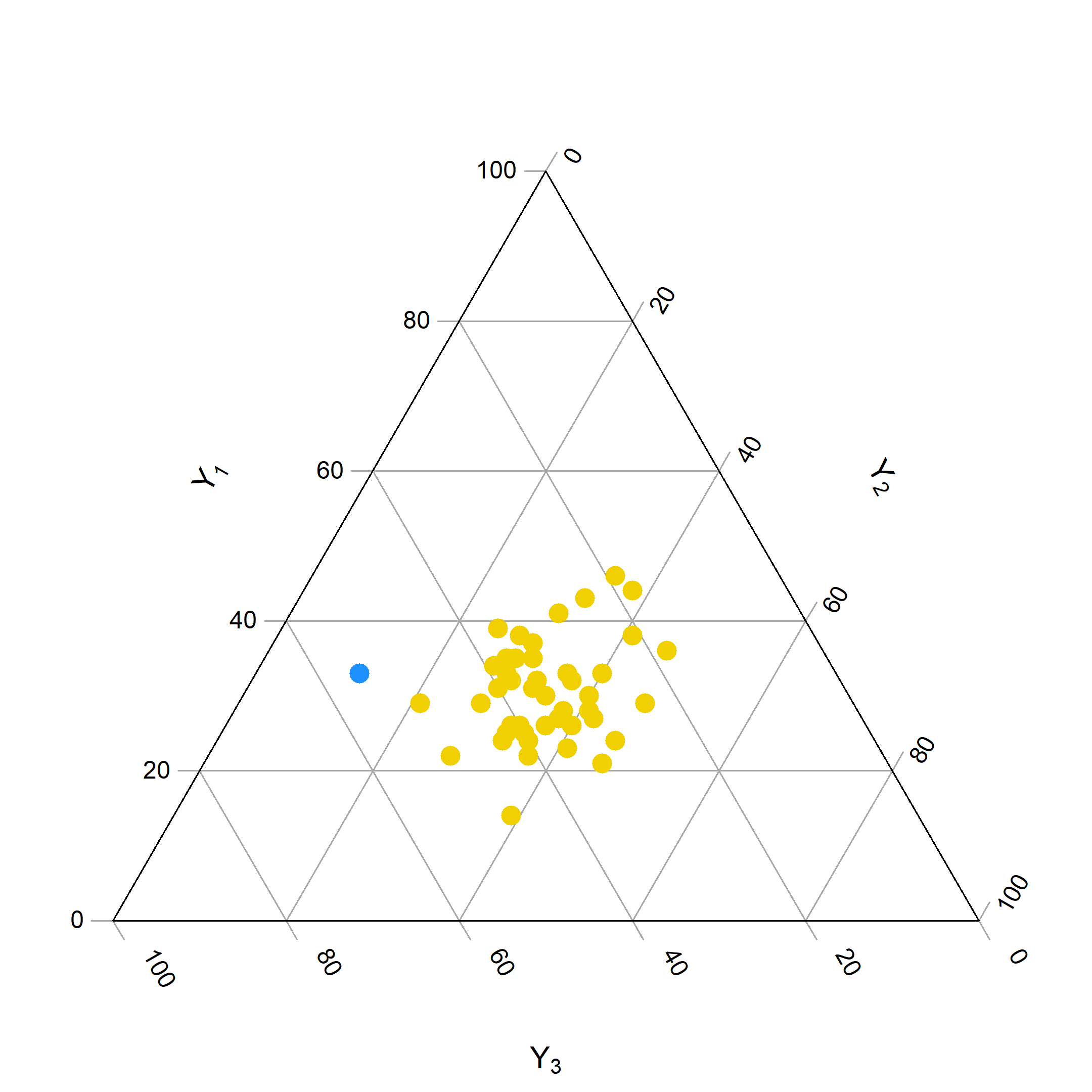}  \label{fig:L_M1}
            \subcaption{Location A: $\by=(33,12,55)$}
	    \end{subfigure}%
        \begin{subfigure}{.5\textwidth} 
		    \centering
            \includegraphics[scale=0.55]{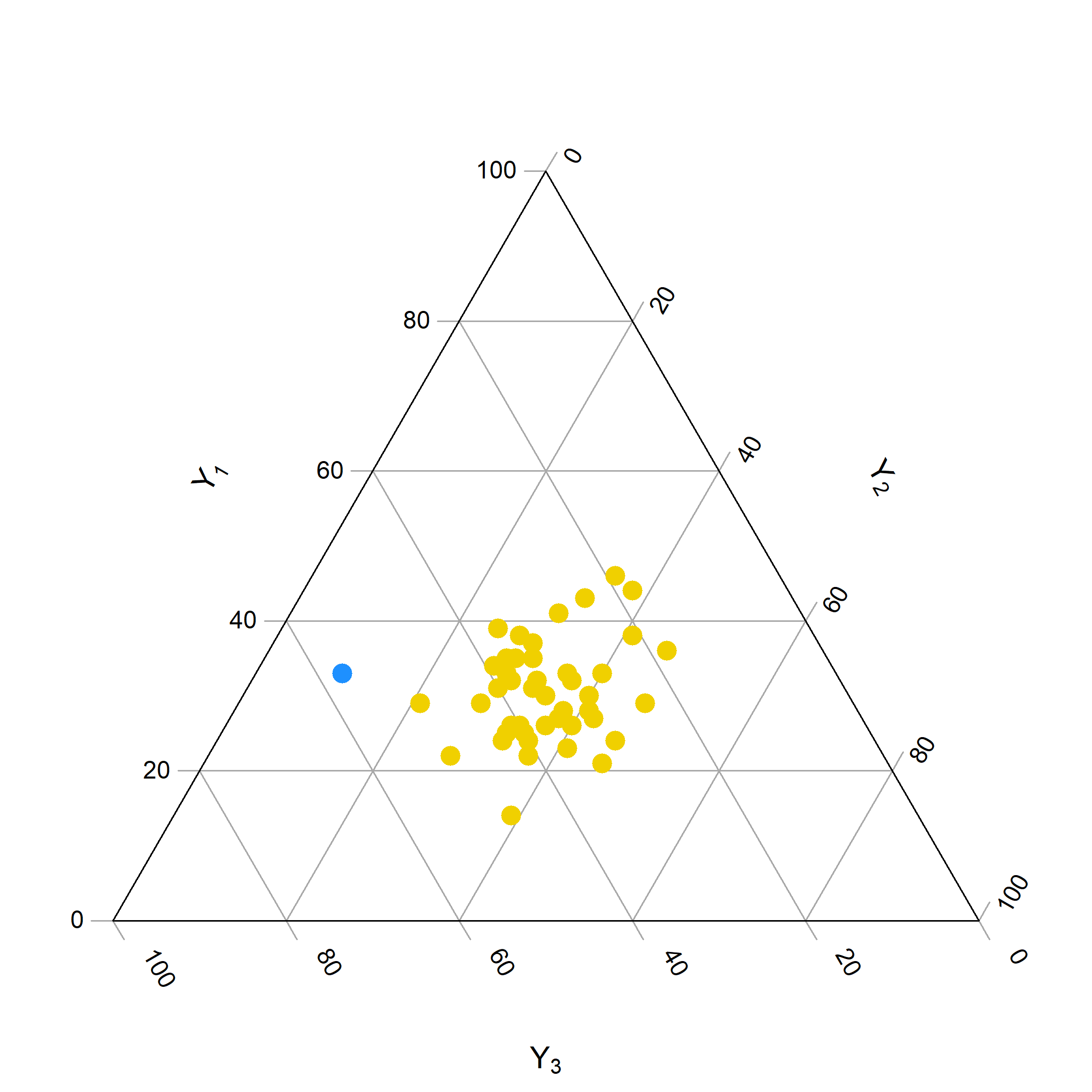}  \label{fig:L_M2}
            \subcaption{Location B: $\by=(33,10,57)$}
	    \end{subfigure}
        \begin{subfigure}{.5\textwidth}
		    \centering
            \includegraphics[scale=0.55]{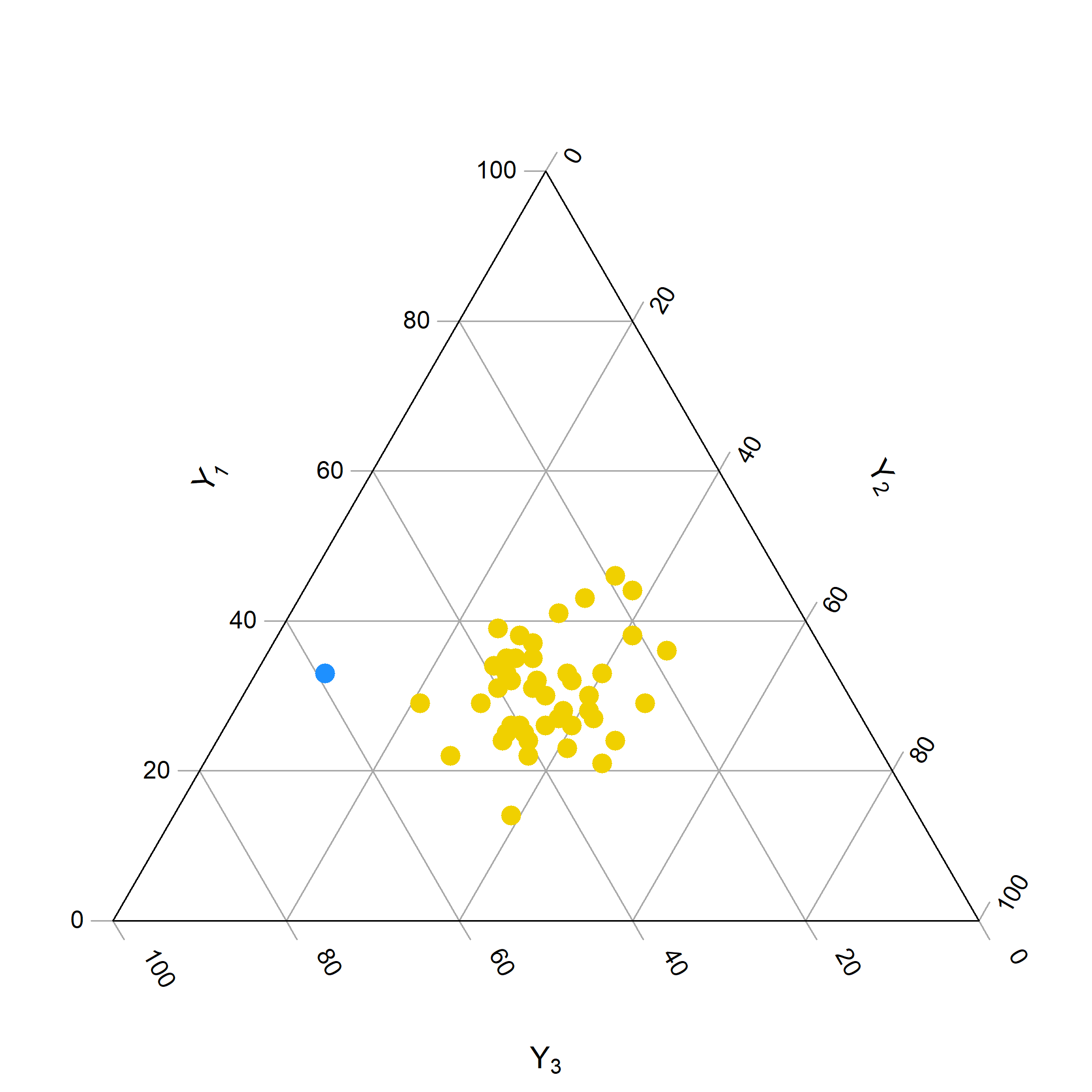}  \label{fig:L_M3}
            \subcaption{Location C: $\by=(33,8,59)$}
	    \end{subfigure}%
        \begin{subfigure}{.5\textwidth} 
		    \centering
            \includegraphics[scale=0.55]{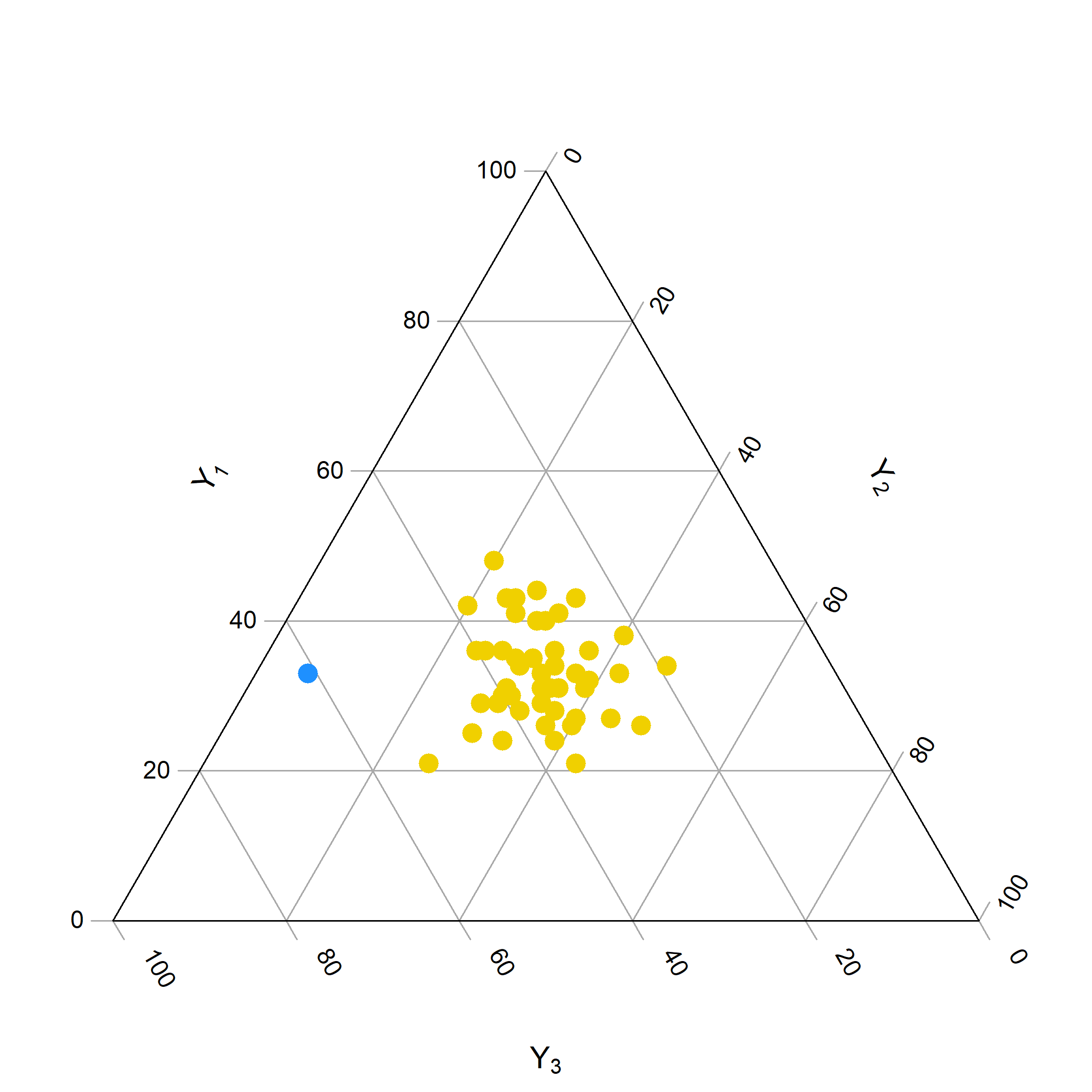}  \label{fig:L_M4}
            \subcaption{Location D: $\by=(33,6,61)$}
	    \end{subfigure}%
        \caption{Four simulated datasets from Scenario 2 of the sensitivity analysis (impact of a single anomalous observation), corresponding to the four anomaly locations (A--D). 
        The anomalous point is shown in blue.}
        \label{fig:L_M}
    \end{figure}

In this study, we generate 49 observations from the DM distribution in \eqref{eq:DM_PPV}, with dimension $D = 3$ and total count $m = 100$, using parameters $\bm{\pi} = (1/3, 1/3)$ and $\sigma = 0.01$. 
Two distinct scenarios (Scenario~1 and Scenario~2) are considered, each differing in the direction along which a single anomalous observation is introduced.
Within each scenario, we artificially perturb the data by introducing a single anomalous point, yielding a sample of size $n=50$. 
The anomalous point is placed at four distinct locations---labeled A, B, C, and D---such that it progressively moves farther from the main data cluster. 
This design produces a total of $2 \times 4 = 8$ unique data configurations. Examples of the two scenarios are illustrated in Figures~\ref{fig:L_C} and~\ref{fig:L_M} using ternary plots.




\begin{figure}[!ht] 
    \centering
    \includegraphics[scale=0.45]{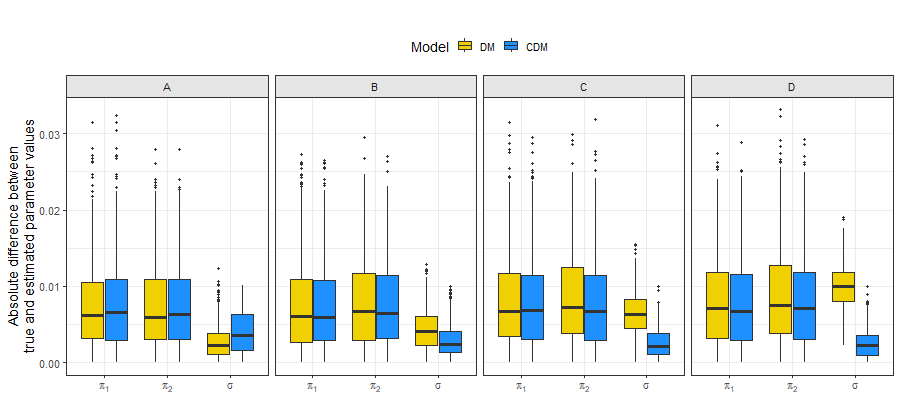} 
    \caption{Boxplots of the absolute differences between the true and estimated parameter values for Scenario~1 across the four locations (A--D).}
    \label{fig:L_C_Box} 
\end{figure}

\begin{figure}[!ht] 
    \centering
    \includegraphics[scale=0.45]{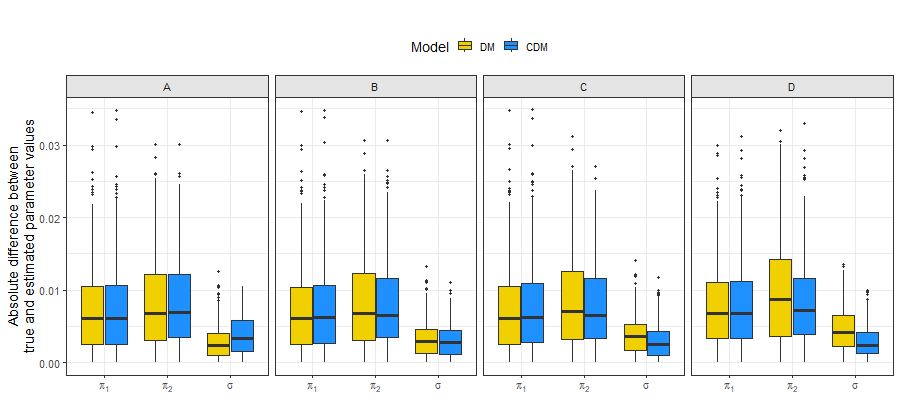} 
    \caption{Boxplots of the absolute differences between the true and estimated parameter values for Scenario~2 across the four locations (A--D).}
    \label{fig:L_M_Box} 
\end{figure}

For each of the eight data configurations, 500 datasets are simulated, yielding a total of $8 \times 500 = 4000$ datasets. Both the DM and CDM distributions are fitted to each simulated dataset, and the absolute differences between the true and estimated parameter values of $\bm{\pi}$ and $\sigma$ are computed for each replication. 
The results are summarized using boxplots: \figurename~\ref{fig:L_C_Box} corresponds to Scenario~1 and \figurename~\ref{fig:L_M_Box} to Scenario~2, with each panel displaying outcomes across the four locations (A--D).

A key observation is that the estimation errors for the CDM distribution remain largely stable across all data configurations. 
In contrast, the performance of the DM distribution deteriorates progressively within each scenario as the anomalous observation shifts from location A to D, moving farther from the bulk of the DM-generated data. 
Only at locations A and B do the parameter estimates of the CDM distribution exhibit higher mean squared error, likely because the anomalous observation lies close to the main data cluster—an intuitively expected outcome.

This behaviour is further supported by the classification results in Tables~\ref{tab:L_Rates_C} and~\ref{tab:L_Rates_M}. 
In both scenarios, the true positive rate (TPR) increases as the anomalous observation moves away from the bulk of the data, approaching one at the more extreme locations. 
This indicates that the CDM model becomes increasingly effective at identifying anomalies as their separation from the main cluster grows. 
Conversely, at locations A and B, where the anomalous observation is less distinct, the TPR is lower, reflecting the greater difficulty in distinguishing such points from typical observations. 
Across all configurations, the false positive rate (FPR) remains low, demonstrating that the CDM model rarely misclassifies typical observations as anomalous.

\begin{table}[!ht]
	\caption{Classification results for Scenario~1 of the sensitivity analysis (impact of a single anomalous observation) under the CDM distribution.}
	\centering
    \scalebox{0.9}{
	\begin{tabular}{@{}lrrrr@{}}
		\toprule
		                                & A: $\by=(20,20,60)$ &B: $\by=(16,16,68)$ &C: $\by=(12,12,76)$ &D: $\by=(8,8,84)$    \\ \midrule
		    True Positive Rate          & 0.550               & 0.960              & 1.000              & 1.000               \\
		    False Positive Rate         & 0.085               & 0.017              & 0.006              & 0.004               \\ \bottomrule
	\end{tabular}} \label{tab:L_Rates_C}
\end{table}

\begin{table}[!ht]
	\caption{Classification results for Scenario~2 of the sensitivity analysis (impact of a single anomalous observation) under the CDM distribution.}
	\centering
    \scalebox{0.9}{
	\begin{tabular}{@{}lrrrr@{}}
		\toprule
		                                 & A: $\by=(33,12,55)$ & B: $\by=(33,10,57)$ & C: $\by=(33,8,59)$ & D: $\by=(33,6,61)$   \\ \midrule
		    True Positive Rate        & 0.668               & 0.868               & 0.924              & 0.962                \\
		    False Positive Rate       & 0.063               & 0.035               & 0.018              & 0.010                \\ \bottomrule
	\end{tabular}} \label{tab:L_Rates_M}
\end{table}


\subsection{Sensitivity analysis: background noise} \label{subsec:variance}

In this study, we simulate 100 datasets of size 900 from the DM distribution in \eqref{eq:DM_PPV}, with $D = 3$, $m = 3$, and $\sigma = 0.01$, under two different parameter configurations for $\bm{\pi}$. These are referred to as Scenario~1, where $\bm{\pi} = \left(\frac{1}{3}, \frac{1}{3}\right)$, and Scenario~2, where $\bm{\pi} = \left(\frac{1}{4}, \frac{1}{4}\right)$. 
In both cases, an additional 100 observations are generated from a uniform distribution over the interval $[0,1]$, resulting in datasets of size $n = 1000$. 
Each noisy observation is then rescaled to satisfy the unit-sum constraint.

\begin{figure}[!ht] 
    \centering
    \includegraphics[scale=0.45]{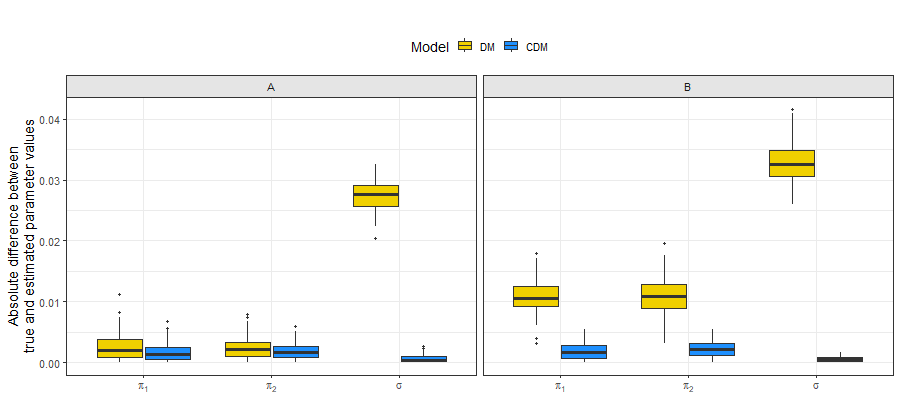} 
    \caption{Boxplots of the absolute differences between the true and estimated parameter values for the sensitivity analysis (background noise), with Scenario~1 shown in the left panel and Scenario~2 in the right panel.}
    \label{fig:L_M_Box var} 
\end{figure}

\begin{figure}[!ht] 
	\centering
	\begin{subfigure}{.5\textwidth}
		\centering
        \includegraphics[scale=0.55]{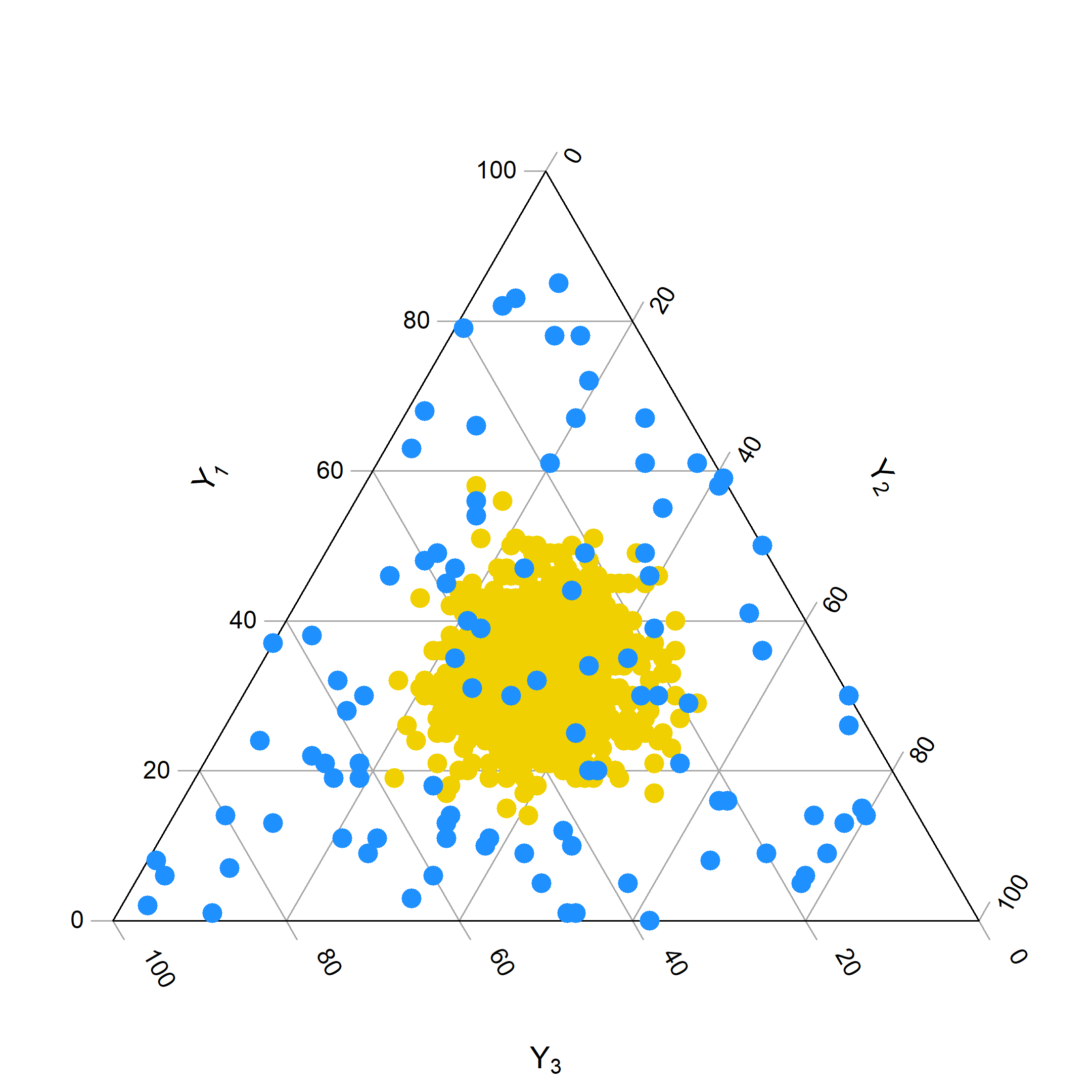}  
        \subcaption{Generated data (yellow) and noise (blue).}
	\end{subfigure}%
    \begin{subfigure}{.5\textwidth} 
		\centering
        \includegraphics[scale=0.55]{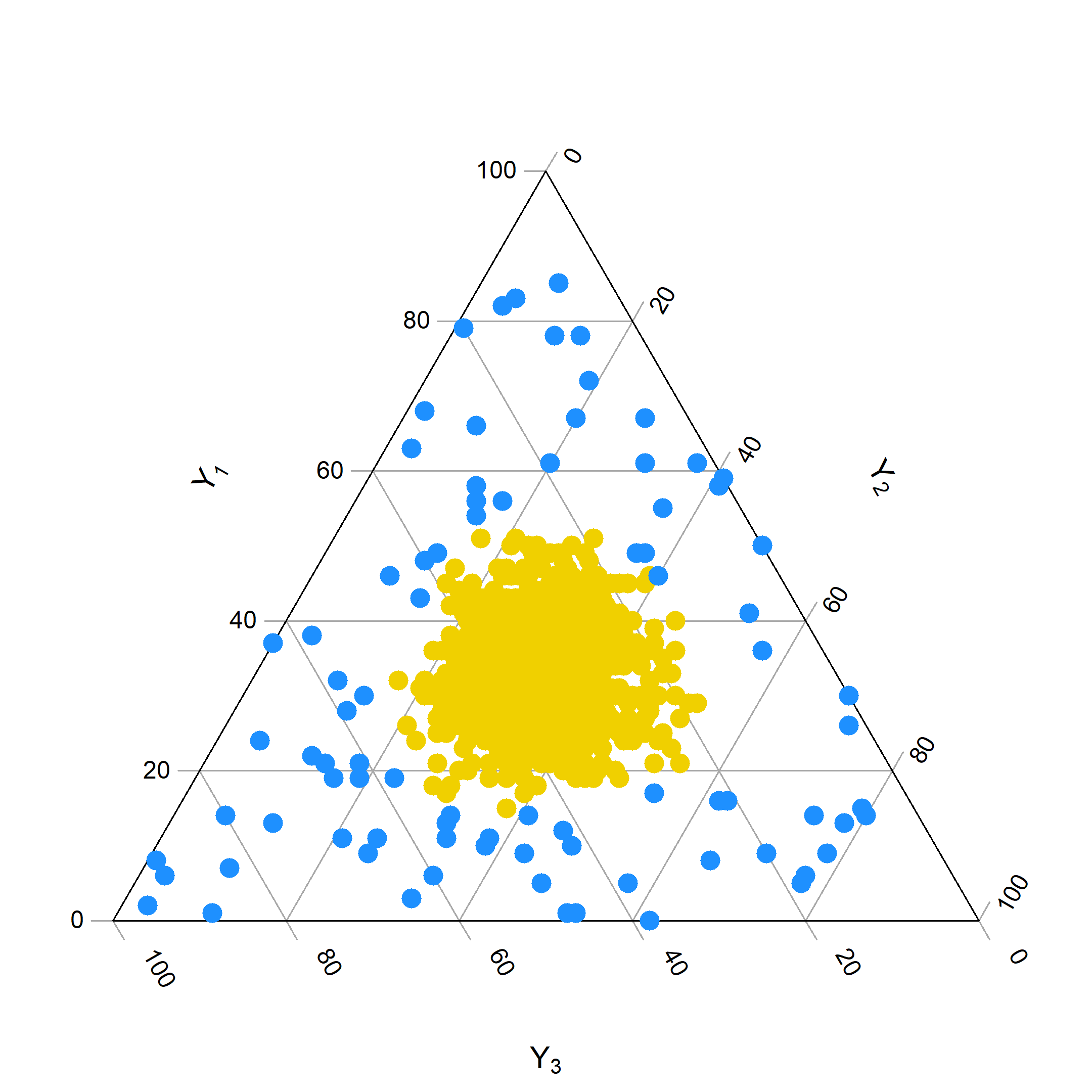}  \label{fig:V_333_class}
        \subcaption{CDM classification: non-outliers (yellow) and outliers (blue).}
	\end{subfigure}%
    \caption{Data generated from a DM distribution with parameters $\bm{\pi} = \left(\frac{1}{3}, \frac{1}{3}\right)$ and $\sigma = 0.01$.}
    \label{fig:V_333}
\end{figure}

\begin{figure}[!ht] 
	\centering
	\begin{subfigure}{.5\textwidth}
		\centering
        \includegraphics[scale=0.55]{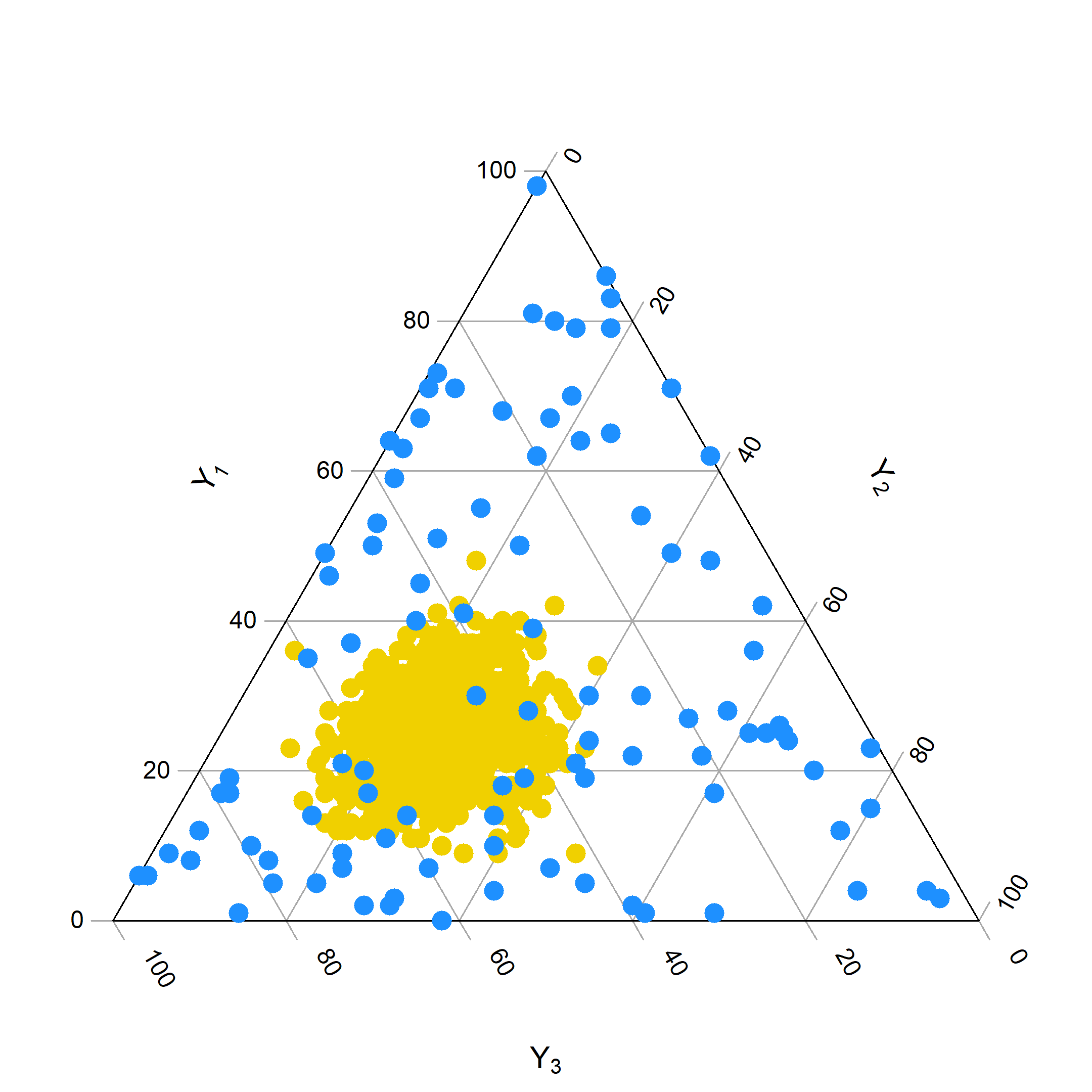}  \l
        \subcaption{Generated data (yellow) and noise (blue).}
	\end{subfigure}%
    \begin{subfigure}{.5\textwidth} 
		\centering
        \includegraphics[scale=0.55]{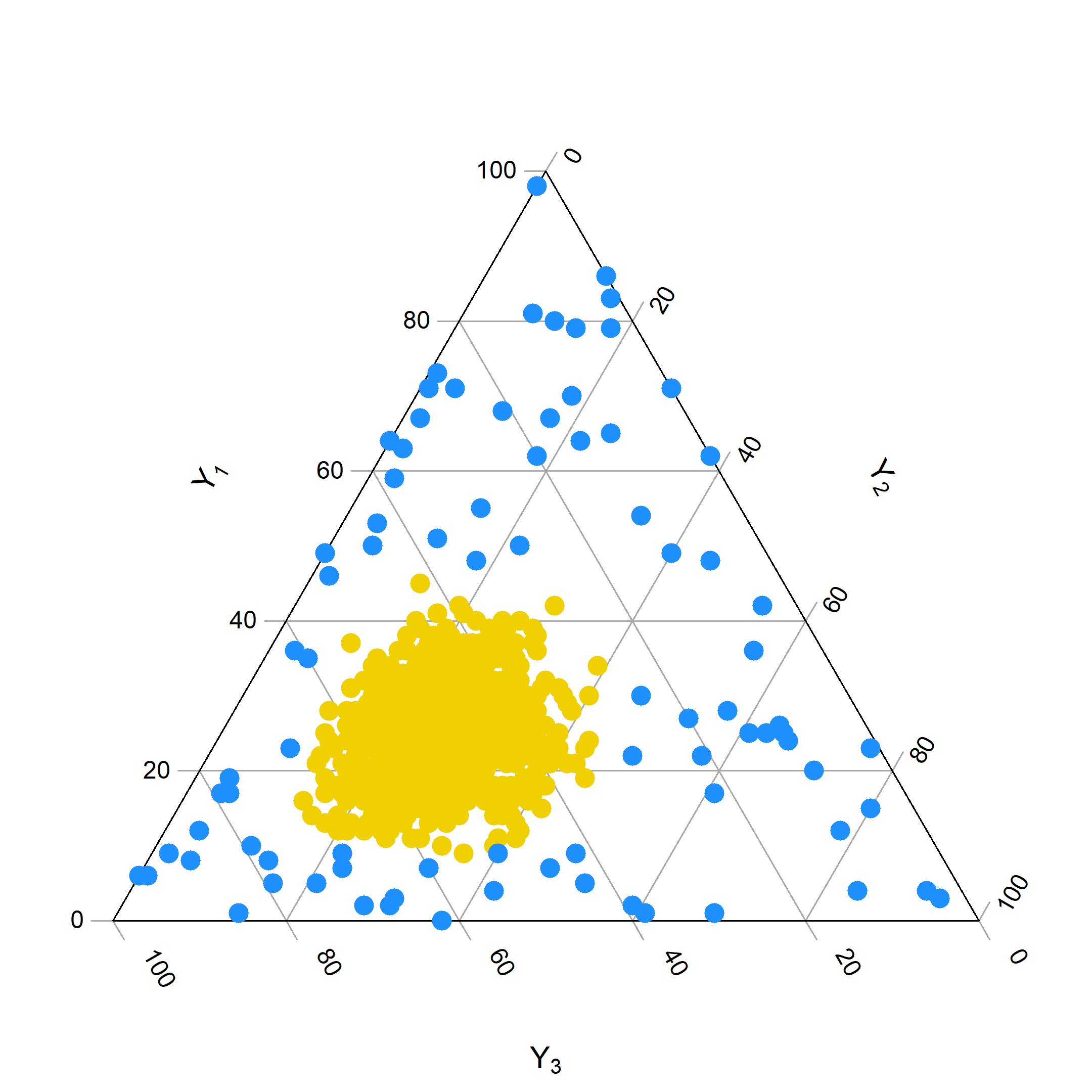}  \label{fig:V_623_class}
        \subcaption{CDM classification: non-outliers (yellow) and outliers (blue).}
	\end{subfigure}%
    \caption{Data generated from a DM distribution with parameters $\bm{\pi} = \left(\frac{1}{4}, \frac{1}{4}\right)$ and $\sigma = 0.01$.}
    \label{fig:V_623}
\end{figure}

The datasets generated under Scenarios 1 and 2 are displayed in \figurename~\ref{fig:V_333}(a) and \figurename~\ref{fig:V_623}(a), respectively. Here, the noisy observations are shown in blue. The corresponding classifications obtained from the CDM model, based on the a posteriori probability in \eqref{eq:posterior_prob}, are presented in \figurename~\ref{fig:V_333}(b) and \figurename~\ref{fig:V_623}(b). 
In these plots, observations classified as good are depicted in yellow, whereas detected anomalies are shown in blue. The results are summarized via box-plots, in \figurename~\ref{fig:L_M_Box var}. The higher number of anomalous observations (relative to that in Section \ref{subsec:location}), highlights the necessity of the CDM distribution.

The poor performance of the DM distribution in Scenario 2 of this sensitivity analysis is motivated as follows: In Scenario 1, the mean of the noisy component lies at the centre of the domain, aligning with the center of the generated data. 
Consequently, the estimated mean proportions $\bpi$ remain largely unaffected by the presence of noisy observations.
In contrast, in Scenario 2, the mean proportion vector is displaced from the center of the domain. As a result, the noisy observations introduce a significant bias in estimating the mean proportion vector. 
This again highlights the importance of the CDM model.

The lack of convergence of the TPR toward one does not necessarily indicate an error. Because of the way the background noise is incorporated into the data, some of them may resemble typical observations (see Figures \ref{fig:V_333} and \ref{fig:V_623}), leading the CDM distribution to classify them as typical. 
This overlap lowers the TPR to 0.749 for Scenario 1 and 0.778 for Scenario 2. In contrast, the FPR remains consistently low -- averaging below 0.01 for both studies (see \tablename~\ref{tab:V_Rates})—demonstrating that the CDM distribution rarely misclassifies typical observations as anomalies



\begin{table}[!ht]
	\caption{True and false positive rates for the sensitivity analysis (background noise)}
	\centering
    \scalebox{0.9}{
	\begin{tabular}{@{}lrr@{}}
		\toprule
		                                 &Scenario 1: $\bm{\pi}=(\frac{1}{3},\frac{1}{3})$ &Scenario 2: $\bm{\pi}=(\frac{1}{4},\frac{1}{4})$  \\  \midrule
		    True Positive Rate        & 0.749                                      & 0.778                                       \\
		    False Positive Rate       & 0.005                                      & 0.004                                       \\  \midrule
	\end{tabular}} \label{tab:V_Rates}
\end{table}

\clearpage

\section{Data Application}
\label{sec:Data Application}



The data considered here originate from \citet{nakatsu2015gut}, who conducted 16S rRNA sequencing on mucosal biopsy samples collected across different stages of colorectal carcinogenesis. 
The dataset was previously analysed by \cite{Subedi}, who focused on the five most abundant bacterial phyla: Firmicutes, Proteobacteria, Bacteroidetes, Fusobacteria, and Actinobacteria. 
An additional category was constructed by aggregating the remaining taxa, yielding a total of $D = 6$ count components.

In this study, we restrict attention to two subsets of the data: (i) samples from healthy control subjects, hereafter referred to as the healthy subset, and (ii) samples from tissue adjacent to tumours in colon cancer patients, hereafter referred to as the carcinoma subset. For each observation, the response is a vector of $D = 6$ count components.
Importantly, the total count varies across samples, a feature that is naturally accommodated by both the DM and CDM models.

For each subset, the multinomial (M) distribution is also fitted to the data via maximum likelihood, alongside the DM and CDM models, using the \textbf{MGLM} package \citep{MGLMpack,MGLMpap}.

\subsection{Healthy subset}
\label{subsec:Healthy subset}

\begin{figure}[!ht] 
    \centering
    \includegraphics[scale=0.4]{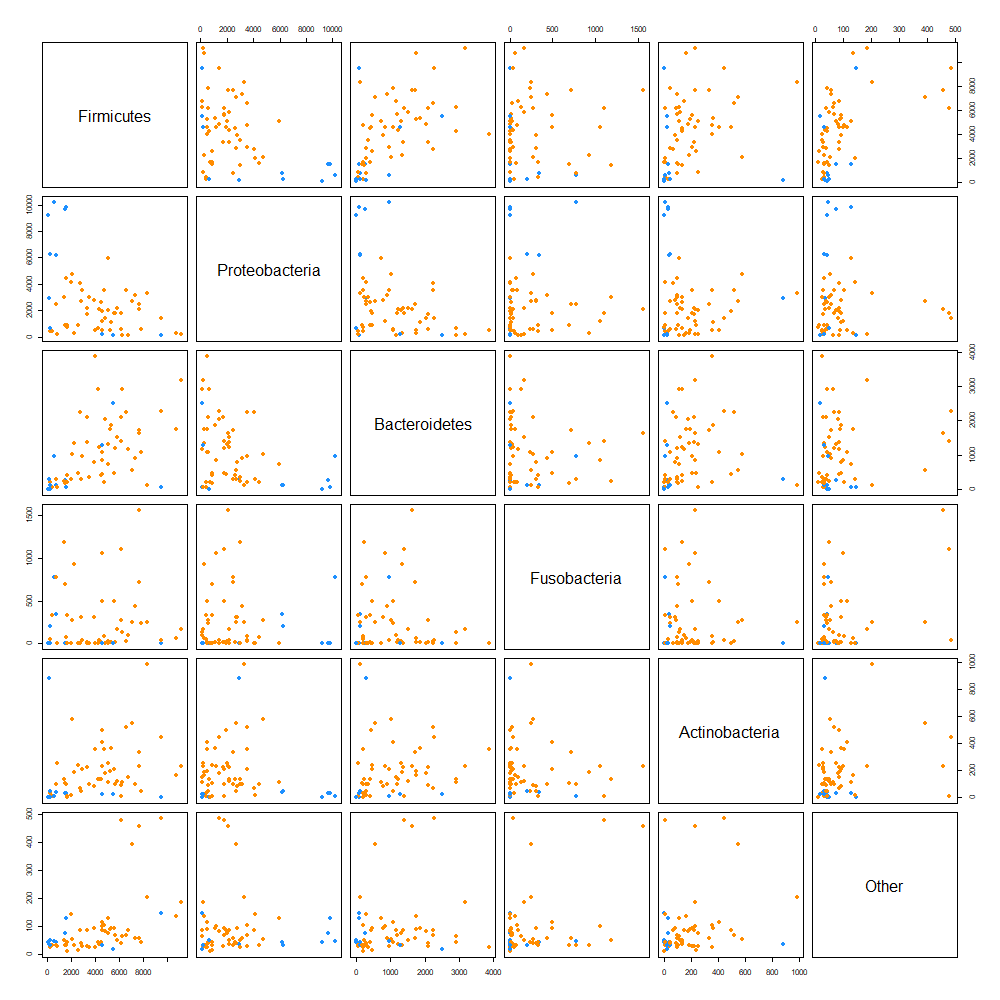} 
    \caption{Pairwise scatter plot matrix for the healthy dataset, with colour indicating typical (orange) and anomalous (blue) observations.}
    \label{fig:L_623} 
\end{figure}

The healthy subset comprises 61 observations. Parameter estimates for the competing models on this subset, together with log-likelihood (LL), AIC, and BIC, are presented in \tablename~\ref{tab:Healthy_Parameter}.
The estimated mean proportions under the DM distribution are distributed across the five taxa, with Firmicutes receiving the largest share ($\hat \pi_1 = 0.430$) and Actinobacteria the smallest ($\hat \pi_5 = 0.065$), and a dispersion estimate of $\hat \sigma = 0.174$. Under the CDM distribution, the estimated mean proportions shift noticeably---particularly $\hat \pi_2$ through $\hat \pi_5$, which are markedly smaller---while $\hat \sigma$ reduces substantially to $0.091$. 
This decrease in $\hat{\sigma}$ indicates that, once anomalous observations are accounted for by the contaminant component of the CDM distribution, the remaining reference component exhibits substantially lower dispersion. 
The estimated contamination proportion ($\hat{\delta} = 0.213$) implies that approximately $21.3\%$ of the observations are attributed to the contaminant component, with a corresponding degree of contamination of $\hat{\eta} = 4.009$. 
This suggests the presence of a notable fraction of observations that deviate from the reference distribution and exhibit markedly higher variability. 
Such heterogeneity in the microbiome profiles may reflect underlying biological variability or experimental noise. 
Of the 61 observations, 11 were identified as anomalous according to the posterior classification rule in \eqref{eq:posterior_prob}, as shown in \figurename~\ref{fig:L_623}, where anomalies are indicated in blue.
This separation is also visually supported by \figurename~\ref{fig:L_623}, where the observations identified as anomalous tend to lie at the periphery of the data cloud.

\begin{table}[!ht]
    \caption{Estimated parameter values and model evaluation metrics---log-likelihood (LL), AIC, and BIC---for the fitted models on the healthy subset.}
    \centering
    \scalebox{0.9}{
    \begin{tabular}{@{}lrrrrrrrrrrr@{}}
        \toprule
                & $\hat \pi_1$  & $\hat \pi_2$  & $\hat \pi_3$  & $\hat \pi_4$  & $\hat \pi_5$  & $\hat \sigma$ & $\hat \eta$ & $\hat \delta$& LL         & AIC        & BIC         \\ \midrule
            M   & 0.483       & 0.317       & 0.125       & 0.036       & 0.026       & -             & -           & -         & -128328.700   & 256667.500   &  256678.000 \\
            DM  & 0.430       & 0.262       & 0.146       & 0.038       & 0.065       & 0.174         & -           & -         &   -2136.506 &   4285.013 &    4297.678 \\
            CDM & 0.464       & 0.218       & 0.137       & 0.030       & 0.053       & 0.091         & 4.009       & 0.213     &  -2120.767 &   4257.534 &    4274.421 \\ \bottomrule
    \end{tabular}} 
    \label{tab:Healthy_Parameter}
\end{table}

The CDM distribution achieves a lower AIC ($4257.534$ vs.\ $4285.013$) and BIC ($4274.421$ vs.\ $4297.678$) than the DM distribution, demonstrating a superior fit.
The improvement in both AIC and BIC is substantial, indicating that the additional flexibility of the CDM model is justified despite the increased number of parameters. 
This suggests that the presence of anomalous observations is substantial enough to warrant a contaminated modelling framework rather than a standard DM specification.

\clearpage

\subsection{Carcinoma subset}
\label{subsec:Carcinoma subset}

The carcinoma subset consists of 52 observations. 
The estimated parameters for the competing models, along with the corresponding log-likelihood, AIC and BIC values, are reported in \tablename~\ref{tab:Cancer_Parameter}.

\begin{table}[!ht]
    \caption{Estimated parameter values and evaluation metrics for Carcinoma subset}
    \centering
    \scalebox{0.9}{
    \begin{tabular}{@{}lrrrrrrrrrrr@{}}
        \toprule
                & $\hat \pi_1$& $\hat \pi_2$& $\hat \pi_3$& $\hat \pi_4$& $\hat \pi_5$& $\hat \sigma$ & $\hat \eta$ & $\hat \delta$ & LL         & AIC        & BIC         \\ \midrule
            M   & 0.542       & 0.151       & 0.153       & 0.105       & 0.020       & -             & -           & -             & -107882.900    & 215775.800   & 215785.600    \\
            DM  & 0.519       & 0.123       & 0.171       & 0.079       & 0.044       & 0.125         & -           & -             &   -1897.525  &   3807.049 &    3818.757 \\
            CDM & 0.547       & 0.102       & 0.165       & 0.065       & 0.034       & 0.068         & 4.116       & 0.178         &   -1883.845  &   3783.691 &    3799.301 \\ \bottomrule
    \end{tabular}} \label{tab:Cancer_Parameter}
\end{table}

\begin{figure}[!ht]
    \centering
    \includegraphics[scale=0.4]{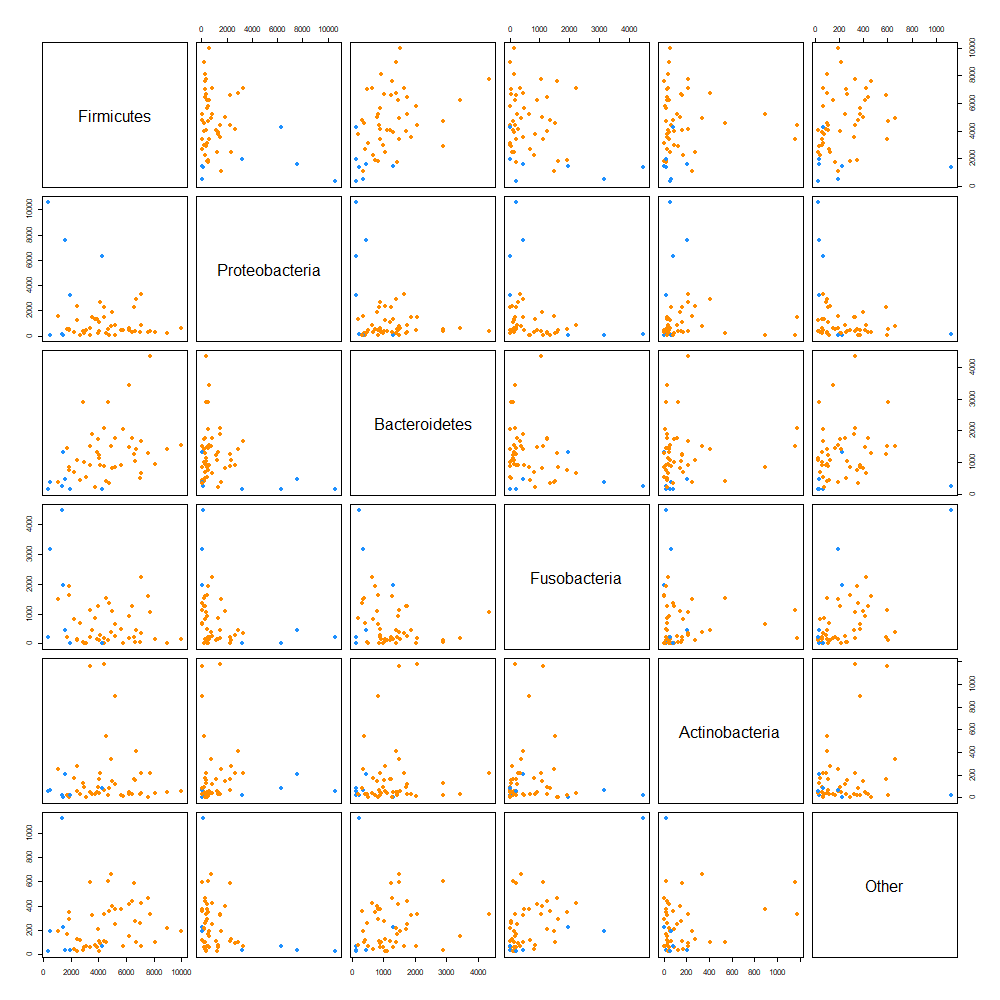}
    \caption{Pairwise scatter plot matrix for the Carcinoma dataset with colour indicating the typical (in orange) and anomalous observations (in blue).}
    \label{fig:Cancer_Pairs}
\end{figure}

Under the DM distribution, Firmicutes again dominates with $\hat \pi_1 = 0.519$, while the remaining taxa receive moderate shares, and $\hat \sigma = 0.125$ indicates moderate dispersion. 
The inflation factor $\hat \eta = 4.116$ indicates that the anomalous observations are considerably more dispersed than those from the reference component, and the mixing proportion $\hat{\delta} = 0.178$ implies that approximately $17.8\%$ of observations are attributed to this component. 
Compared to the healthy subset in Section~\ref{subsec:Healthy subset}, the estimated contamination proportion is slightly lower, suggesting a somewhat reduced presence of atypical observations. As in the healthy subset, the reduction in the estimated dispersion parameter under the CDM model indicates that part of the variability captured by the DM model is driven by anomalous observations rather than genuine overdispersion. 
However, the contamination remains substantial, indicating that heterogeneity is also a relevant feature in the carcinoma subset. 
Of the 52 observations, 7 were identified as anomalous according to the posterior classification rule. 
As illustrated in \figurename~\ref{fig:Cancer_Pairs}, these observations tend to lie at the periphery of the data cloud, suggesting that the detected anomalies correspond to observations with atypical compositions rather than random noise. 
The separation between typical and anomalous observations is particularly evident in several pairwise projections, further supporting the ability of the CDM model to capture underlying heterogeneity in the data. As shown in \tablename~\ref{tab:Cancer_Parameter}, the CDM model again outperforms the DM across all considered metrics.

\clearpage

\section{Conclusion} 
\label{section:conclusion}

The human gut microbiome plays a central role in human health, and its accurate statistical characterisation has direct consequences for biological inference and clinical understanding. Disruptions to microbial equilibrium have been implicated in colorectal cancer \citep{nakatsu2015gut}, inflammatory bowel disease \citep{halfvarson2017dynamics}, and a broad range of systemic conditions \citep{lloyd2019multi}, making robust modelling of microbiome composition data a pressing methodological priority. 
The Dirichlet-multinomial (DM) distribution is widely adopted for such compositional count data, as it accommodates overdispersion beyond the restrictive multinomial assumption. 
However, the DM distribution itself can prove insufficiently flexible when anomalous observations disproportionately inflate the estimated dispersion, potentially leading to misleading biological conclusions.

We propose the contaminated Dirichlet–multinomial (CDM) distribution, a two-component mixture in which observations arise either from a reference DM distribution or from a DM distribution with an inflated dispersion parameter. 
Both components share the same mean proportion vector, so that differences between them are captured solely through their dispersion.
The CDM distribution thereby protects the reference distribution from the influence of atypical observations, and the contamination parameters $\delta$ and $\eta$ have direct interpretations as the proportion and severity of anomalous observations, respectively. 
A posterior classification rule enables automatic outlier detection without requiring pre-specification of suspect observations. The CDM distribution also simplifies to the contaminated beta-binomial distribution proposed in \citet{otto2026modeling}.

Two sensitivity analyses confirm the robustness of the CDM distribution against both isolated and diffuse contamination. 
In particular, the estimation error of the DM model deteriorates markedly in the presence of anomalies, whereas the CDM distribution remains stable.

When applied to 16S rRNA sequencing data from \cite{nakatsu2015gut}, the CDM model consistently outperforms both the DM and the nested multinomial model across the healthy and carcinoma subsets, as measured by classical model selection criteria such as AIC and BIC. 
In addition, it identifies 11 and 7 atypical observations, respectively, while yielding substantially reduced dispersion estimates for the reference component.

These results highlight the value of the CDM distribution as a robust and interpretable tool for microbiome composition analysis, where the ability to distinguish genuine biological signal from anomalous observations is critical for reliable downstream inference \citep{lloyd2019multi, halfvarson2017dynamics}. Although the model is introduced and motivated in the context of microbiome data, its applicability extends more broadly to settings involving multinomial count data, where contamination and excess variability are common features of real-world datasets.

 \section*{Acknowledgements}
 This work was based upon research supported in part by the National Research Foundation (NRF) of South Africa (SA), grant RA231117164450, the Centre of Excellence in Mathematical and Statistical Sciences, based at the University of the Witwatersrand (SA). The opinions expressed and conclusions arrived at are those of the authors and are not necessarily to be attributed to the NRF. 

\section*{Data availability statement}
All datasets considered in this paper are freely available on the internet.

\section*{Disclosure statement}
The authors declared no potential conflicts of interest with respect to the research, authorship, and/or publication of this article.

\appendix

\section{A practitioners guide to the \textbf{CDM} package}\label{App:CDM package}
\definecolor{bg}{rgb}{.9, .9, .9}

This appendix serves as a ``practitioner’s guide'' to implementing the methodology presented in this paper. It provides details on the \textbf{CDM} package for \textsf{R}, along with examples that reproduce the results in Section \ref{subsec:Healthy subset}, and illustrate its broader application. The package is available on GitHub at \url{https://github.com/u20439530/CDM_Package}
.
\subsection*{Installation}

To install the \textbf{CDM} package directly from GitHub, use the following code in \textsf{R}:
\begin{Rlst}
#install.packages("devtools")
library(devtools)
install_github("u20439530/CDM_Package")
\end{Rlst}
\subsection*{Main function: \texttt{CDM.fit()}}

The \texttt{CDM.fit()} function performs ML estimation of the CDM distribution in \eqref{eq:CDM_PMF}. Below is a detailed description of its arguments and return values:
\begin{Rlst}
CDM.fit(Y)
\end{Rlst}
\subsubsection*{Arguments}

\begin{longtable}{p{4cm} p{11cm}}
\texttt{Y} & A matrix of count data.
\end{longtable}
\subsubsection*{Return Values}
The \texttt{CDM.fit()} function returns a list with the following components:
\begin{longtable}{p{4cm} p{11cm}}
\texttt{pi} &  Maximum likelihood estimates of the proportional mean ($\pi$). \\
\texttt{sigma} & Maximum likelihood estimates of the dispersion parameter ($\sigma$). \\
\texttt{eta} & Maximum likelihood estimates of inflation parameter ($\eta$). \\
\texttt{delta} & Maximum likelihood estimates of the proportion of observations belonging to the contaminant ($\delta$). \\
\texttt{LL} & The log-likelihood value at convergence. \\
\texttt{AIC} & Akaike Information Criterion. \\
\texttt{BIC} & Bayesian Information Criterion. \\
\texttt{Probability.Posterior} & a posteriori probability that an observation comes from the Dirichlet Multinomial component. \\
\end{longtable}

\subsubsection*{Example}
The following code reproduces the DM and CDM fit in the healthy control subjects Section \ref{subsec:Healthy subset}.
\begin{Rlst}
library(CDM)
data("Healthy_Subset")
data <- Healthy_Subset
est_DM <- DM.fit(data)
est_CDM <- CDM.fit(data)
est_DM$AIC #returns AIC value of fitted DM distribution
est_DM$BIC #returns BIC value of fitted DM distribution
est_CDM$AIC #returns AIC value of fitted CDM distribution
est_CDM$BIC #returns BIC value of fitted CDM distribution

# Anomaly detection
prob <- est_CDM$Probability.Posterior
ifelse(prob[,1]>0.5, "good", "anomaly")
\end{Rlst}

\subsection*{Other functions in the \textbf{CDM} package}
The \textbf{CDM} package also includes the following functions:

\begin{longtable}{p{4cm} p{11cm}}
\texttt{DM.PMF()} & Computes the PMF of the mean parameterised DM distribution.\\
\texttt{CDM.PMF()} & Computes the PMF of the CDM distribution. \\
\texttt{DM.fit()} & ML estimation of the mean parameterised DM distribution. \\
\end{longtable}
\noindent For more details, use the \texttt{help} command or by typing \texttt{?function\_name} in \textsf{R} (e.g., {R}{?CDM.fit}).

\bibliographystyle{chicago}
\bibliography{database.bib}

@article{Subedi,
    Author = {Sanjeena Subedi and Drew Neish and Stephen Bak and Zeny Feng},
	Journal = {Journal of the Royal Statistical Society Series C: Applied Statistics},
	Month = {},
	Number = {5},
	Numpages = {25},
	Pages = {1163--1187},
	Title = {Cluster analysis of microbiome data by using mixtures of {D}irichlet–multinomial regression models},
	Volume = {69},
	Url = {},
	Year = {2020}}

@article{Mosimann_1962,
    author = {James E. Mosimann},
	Journal = {Biometrika},
	Month = {},
	Number = {1/2},
	Numpages = {},
	Pages = {65-82},
	Title = {On the compound multinomial distribution, the multivariate $\beta$-distribution, and correlations among proportions},
	Volume = {49},
	Url = {},
	Year = {1962}
}

@article{Akaike_1974,
    author = {Hirotugu Akaike},
	Journal = {IEEE Transactions on Automatic Control},
	Month = {},
	Number = {6},
	Numpages = {},
	Pages = {716-723},
	Title = {A new look at statistical model identification},
	Volume = {19},
	Url = {},
	Year = {1974}
}

@article{Schwarz_1978,
    author = {Gideon Schwarz},
	Journal = {Institute of Mathematical Statistics},
	Month = {},
	Number = {2},
	Numpages = {},
	Pages = {461-464},
	Title = {Estimating the dimension of a model},
	Volume = {6},
	Url = {},
	Year = {1978}
}

@article{otto2026modeling,
  title={Modeling Bounded Count Environmental Data Using a Contaminated Beta-Binomial Regression Model},
  author={Otto, Arnoldus F and Punzo, Antonio and Ferreira, Johannes T and Bekker, Andri{\"e}tte and Tomarchio, Salvatore D and Tortora, Cristina},
  journal={Environmetrics},
  volume={37},
  number={1},
  pages={e70067},
  year={2026},
  publisher={Wiley Online Library}
}

@article{Tomarchio_2020,
    author = {Salvatore D. Tomarchio and Antonio Punzo},
	Journal = {Journal of Applied Statistics},
	Month = {},
	Number = {13-15},
	Numpages = {},
	Pages = {2328-2353},
	Title = {Dichotomous unimodal compound models: application to the distribution of insurance losses},
	Volume = {47},
	Url = {},
	Year = {2020}
}

@article{aitkin1980mixture,
  title={Mixture models, outliers, and the {EM} algorithm},
  author={Aitkin, Murray and Wilson, Granville Tunnicliffe},
  journal={Technometrics},
  volume={22},
  number={3},
  pages={325--331},
  year={1980},
  publisher={Taylor \& Francis}
}

@article{teicher1963identifiability,
  title={Identifiability of finite mixtures},
  author={Teicher, Henry},
  journal={The Annals of Mathematical Statistics},
  volume={34},
  number={4},
  pages={1265--1269},
  year={1963},
  publisher={Institute of Mathematical Statistics}
}

@book{titterington1985statistical,
  title={Statistical Analysis of Finite Mixture Distributions},
  author={Titterington, D. M. and Smith, A. F. M. and Makov, U. E.},
  year={1985},
  publisher={John Wiley \& Sons},
  address={Chichester}
}

@article{Holmes_2012,
    author = {Ian Holmes and Keith Harris and Christopher Quince},
	Journal = {PloS One},
	Month = {},
	Number = {2},
	Numpages = {},
	Pages = {e30126},
	Title = {Dirichlet multinomial mixtures: generative models for microbial metagenomics},
	Volume = {7},
	Url = {},
	Year = {2012}
}

@article{Otto_2025_NB,
    author = {Arnoldus F. Otto and Johannes T. Ferreira and Salvatore D. Tomarchio and Andriëtte Bekker and Antonio Punzo},
	Journal = {Statistical Methods in Medical Research},
	Month = {},
	Number = {2},
	Numpages = {},
	Pages = {369-389},
	Title = {A contaminated regression model for count health data},
	Volume = {34},
	Url = {},
	Year = {2025}
}

@article{1,
    author = {},
	Journal = {},
	Month = {},
	Number = {},
	Numpages = {},
	Pages = {},
	Title = {},
	Volume = {},
	Url = {},
	Year = {}
}

@Manual{MGLMpack,
  title = {MGLM: Multivariate Response Generalized Linear Models},
  author = {Yiwen Zhang and Hua Zhou},
  year = {2022},
  note = {R package version 0.2.1},
  url = {https://CRAN.R-project.org/package=MGLM},
}

@article{MGLMpap,
  author  = {Yiwen Zhang and Hua Zhou and Jin Zhou and Wei Sun},
  title   = {Regression Models for Multivariate Count Data},
  journal = {Journal of Computational and Graphical Statistics},
  year    = {2017},
  volume  = {26},
  number  = {1},
  pages   = {1-13},
  doi     = {10.1080/10618600.2016.1154063}
}

@article{nguyen2026identifiability,
  title={On the identifiability of {D}irichlet mixture models},
  author={Nguyen, Hien Duy and Gupta, Mayetri},
  journal={arXiv preprint arXiv:2603.21914},
  year={2026}
}

@article{chen2013variable,
  title={Variable selection for sparse {D}irichlet-multinomial regression with an application to microbiome data analysis},
  author={Chen, Jun and Li, Hongzhe},
  journal={The Annals of Applied Statistics},
  volume={7},
  number={1},
  pages={418--442},
  year={2013}
}

@article{wadsworth2017integrative,
  title={An integrative {B}ayesian {D}irichlet-multinomial regression model for the analysis of taxonomic abundances in microbiome data},
  author={Wadsworth, W Duncan and Argiento, Raffaele and Guindani, Michele and Galloway-Pena, Jessica and Shelburne, Samuel A and Vannucci, Marina},
  journal={BMC bioinformatics},
  volume={18},
  number={94},
  pages={},
  year={2017},
  publisher={Springer}
}

@article{harrison2020dirichlet,
  title={Dirichlet-multinomial modelling outperforms alternatives for analysis of microbiome and other ecological count data},
  author={Harrison, Joshua G and Calder, W John and Shastry, Vivaswat and Buerkle, C Alex},
  journal={Molecular Ecology Resources},
  volume={20},
  number={2},
  pages={481--497},
  year={2020},
  publisher={Wiley Online Library}
}

@article{guimaraes2005dirichlet,
  title={Dirichlet-multinomial regression},
  author={Guimaraes, Paulo and Lindrooth, Richard and others},
  journal={Economics Working Paper Archive at WUSTL, Econometrics},
  volume={509001},
  year={2005}
}

@article{hennig2002fixed,
	title={Fixed point clusters for linear regression: computation and comparison},
	author={Hennig, Christian},
	journal={Journal of {C}lassification},
	volume={19},
	number={2},
	pages={249},
	year={2002},
	publisher={Springer Nature BV}
}

@article{davies1993identification,
	title={The identification of multiple outliers},
	author={Davies, Laurie and Gather, Ursula},
	journal={Journal of the {A}merican {S}tatistical {A}ssociation},
	volume={88},
	number={423},
	pages={782--792},
	year={1993},
	publisher={Taylor \& Francis}
}

@book{ritter2014robust,
  title={Robust cluster analysis and variable selection},
  author={Ritter, Gunter},
  year={2014},
  publisher={CRC Press}
}

@book{barnett1994outliers,
  title={Outliers in {S}tatistical {D}ata},
  author={Barnett, Vic and Lewis, Toby},
  volume={3},
  year={1994},
publisher={Wiley Series in Probability and Mathematical Statistics: Applied Probability and Statistics}

}

@article{punzo2016parsimonious,
  title={Parsimonious mixtures of multivariate contaminated normal distributions},
  author={Punzo, Antonio and McNicholas, Paul D},
  journal={Biometrical {J}ournal},
  volume={58},
  number={6},
  pages={1506--1537},
  year={2016},
  publisher={Wiley Online Library}
}

@article{punzo2017robust,
  title={Robust clustering in regression analysis via the contaminated {G}aussian cluster-weighted model},
  author={Punzo, Antonio and McNicholas, Paul D},
  journal={Journal of {C}lassification},
  volume={34},
  pages={249--293},
  year={2017},
  publisher={Springer}
}

@article{mazza2020mixtures,
  title={Mixtures of multivariate contaminated normal regression models},
  author={Mazza, Angelo and Punzo, Antonio},
  journal={Statistical {P}apers},
  volume={61},
  number={2},
  pages={787--822},
  year={2020},
  publisher={Springer}
}

@article{punzo2021multiple,
  title={Multiple scaled contaminated normal distribution and its application in clustering},
  author={Punzo, Antonio and Tortora, Cristina},
  journal={Statistical Modelling},
  volume={21},
  number={4},
  pages={332--358},
  year={2021},
  publisher={SAGE Publications Sage India: New Delhi, India}
}

@Manual{R_core,
    title = {R: A Language and Environment for Statistical Computing},
    author = {{R Core Team}},
    organization = {R Foundation for Statistical Computing},
    address = {Vienna, Austria},
    year = {2023},
    url = {https://www.R-project.org/},
  }

@article{damgaard2018joint,
  title={The joint distribution of pin-point plant cover data: a reparametrized {D}irichlet--multinomial distribution},
  author={Damgaard, Christian},
  journal={arXiv preprint arXiv:1808.04582},
  year={2018}
}

@article{nakatsu2015gut,
  title={Gut mucosal microbiome across stages of colorectal carcinogenesis},
  author={Nakatsu, Geicho and Li, Xiangchun and Zhou, Haokui and Sheng, Jianqiu and Wong, Sunny Hei and Wu, William Ka Kai and Ng, Siew Chien and Tsoi, Ho and Dong, Yujuan and Zhang, Ning and others},
  journal={Nature communications},
  volume={6},
  number={1},
  pages={8727},
  year={2015},
  publisher={Nature Publishing Group UK London}
}

@article{lloyd2019multi,
  title={Multi-omics of the gut microbial ecosystem in inflammatory bowel diseases},
  author={Lloyd-Price, Jason and Arze, Cesar and Ananthakrishnan, Ashwin N and Schirmer, Melanie and Avila-Pacheco, Julian and Poon, Tiffany W and Andrews, Elizabeth and Ajami, Nadim J and Bonham, Kevin S and Brislawn, Colin J and others},
  journal={Nature},
  volume={569},
  number={7758},
  pages={655--662},
  year={2019},
  publisher={Nature Publishing Group UK London}
}

@article{halfvarson2017dynamics,
  title={Dynamics of the human gut microbiome in inflammatory bowel disease},
  author={Halfvarson, Jonas and Brislawn, Colin J and Lamendella, Regina and V{\'a}zquez-Baeza, Yoshiki and Walters, William A and Bramer, Lisa M and D'amato, Mauro and Bonfiglio, Ferdinando and McDonald, Daniel and Gonzalez, Antonio and others},
  journal={Nature microbiology},
  volume={2},
  number={5},
  pages={17004},
  year={2017},
  publisher={Nature Publishing Group}
}

@article{tvedebrink2009overdispersion,
  title={Overdispersion in allelic counts and $\theta$-correction in forensic genetics},
  author={Tvedebrink, Torben},
  journal={Forensic Science International: Genetics Supplement Series},
  volume={2},
  number={1},
  pages={455--457},
  year={2009},
  publisher={Elsevier}
}

@article{melnykov2026use,
  title={On the Use of Contaminated Normal Distributions for Modeling Data Groups with Heavy Tails and Outliers},
  author={Melnykov, Yana},
  journal={Journal of Classification},
  volume={43},
  number={1},
  pages={66--85},
  year={2026},
  publisher={Springer}
}





\end{document}